\documentclass[12pt]{article}

\pdfoutput=1
\usepackage[colorlinks, 
linkcolor={blue!50!black}, 
citecolor={blue!50!black}, 
urlcolor={blue!50!black}]{hyperref} 
\usepackage{setspace}
\usepackage{graphicx} 
\usepackage{{realboxes}}
\graphicspath{{subtex/img/}} 
\usepackage{caption}
\usepackage{subcaption}
\usepackage{comment}
\usepackage[ruled]{algorithm2e}
\usepackage{geometry}
\usepackage{pdflscape}
\usepackage{multirow,array}
\usepackage{textgreek}

\newcommand\Tstrut{\rule{0pt}{2.6ex}}  
\newcommand\Bstrut{\rule[-1.3ex]{0pt}{0pt}} 

\usepackage{booktabs} 
\usepackage{threeparttable} 
\usepackage{tabularx}
\usepackage{tabulary}
\usepackage{amsmath}
\usepackage{yhmath}
\usepackage{amssymb}
\usepackage{soul}
\newcommand{\sym}[1]{\rlap{#1}} 
\usepackage{siunitx}
\usepackage{tikz}

\usepackage{bbm}

\usepackage{natbib}
\usepackage[blocks]{authblk}

\usepackage{hyperref}

\DeclareMathOperator{\argmax}{argmax}   

\usepackage[capitalise,noabbrev]{cleveref}

\newtheorem{lemma}{Lemma}
\newtheorem{definition}{Definition}
\Crefname{hypothesis}{Hypothesis}{Hypotheses}
\Crefname{lemma}{Lemma}{Lemmata}
\newtheorem{proposition}{Proposition}

\newenvironment{proof}[1][Proof]{\noindent \textbf{#1.} }{\rule{0.5em}{0.5em}}

\begin{document}
\newgeometry{top=2cm,bottom=2cm,left=2cm, right=2cm}
	\begin{titlepage}
		\begin{singlespace}
		\title{On the Emergence of Cooperation \\in the Repeated Prisoner's Dilemma} 
		\author{
			\begin{minipage}{0.49\textwidth}
				\centering
				Maximilian Schaefer \\
				Yale University\footnote{Tobin Center for Economic Policy, 30 Hillhouse Avenue, New Haven, 06511 CT. Email: \href{mailto:maximilian.schaefer@yale.edu}{maximilian.schaefer@yale.edu}. The author is also a research affiliate at the University of Bologna. The simulations presented in this study were performed on the Grace cluster of Yale's Center for Research Computing.} 
			\end{minipage}}
		\date{\today}
		\maketitle
		\vspace{-0.5cm}
		\begin{abstract}
			
			Using simulations between pairs of $\epsilon$-greedy q-learners with one-period memory, this article demonstrates that the potential function of the stochastic replicator dynamics \citep{foster1990stochastic} allows it to predict the emergence of error-proof cooperative strategies from the underlying parameters of the repeated prisoner’s dilemma. The observed cooperation rates between q-learners are related to the ratio between the kinetic energy exerted by the polar attractors of the replicator dynamics under the grim trigger strategy. The frontier separating the parameter space conducive to cooperation from the parameter space dominated by defection can be found by setting the kinetic energy ratio equal to a critical value, which is a function of the discount factor, $f(\delta) = \delta/(1-\delta)$, multiplied by a correction term to account for the effect of the algorithms’ exploration probability. The gradient of the share of cooperative strategies at the frontier increases with the distance between the game parameters and the hyperplane that characterizes the incentive compatibility constraint for cooperation under grim trigger. The simulations cover nearly the entire range of positive distances between the game parameters and the hyperplane and rely on learning rates and exploration probabilities in a range between $0.01$ and $0.1.$
			
			Building on a strand of literature from the neurosciences, which suggests that model-free reinforcement learning is useful to understanding human behavior in risky environments, the  article further explores the extent to which the frontier derived for q-learners also explains the emergence of cooperation between humans playing the infinitely repeated prisoner’s dilemma. Using comprehensive metadata from laboratory experiments that analyze human choices in the infinitely repeated prisoner’s dilemma, the cooperation rates between humans are compared to those observed between q-learners under similar conditions. The Pearson correlation coefficients between the cooperation rates observed for humans and those observed for q-learners are consistently above $0.8$. Consistent with the high correlation, the frontier derived from the simulations between q-learners is also found to predict the emergence of cooperation between humans.   
			
		\end{abstract}
		\setcounter{page}{0}
		\thispagestyle{empty}
	\end{singlespace}
	\end{titlepage}
\restoregeometry

\section{Introduction}
\label{sec:intro}
Understanding and enhancing the emergence of cooperative behavior in the repeated prisoner's dilemma is an active area of research across several disciplines. While the computer science literature focuses on developing algorithms that efficiently learn to cooperate \citep{busoniu2008comprehensive, nguyen2020deep, zhang2021multi}, there appears to be limited research aimed at understanding which parameters of the prisoner's dilemma lead to cooperative behavior among algorithms.  

By contrast, characterizing the payoff parameters necessary for cooperative behavior between humans is an important endeavor for economists. Understanding the conditions under which self-interested agents cooperate is important as it could help remedy coordination failure in any situation in which cooperation is socially optimal but not individually rational \citep{blonski2011equilibrium, dal2011evolution,  fudenberg2012slow, dal2018determinants, Bigoni2022}. 

The contribution of this article is twofold. First, it presents the results of a comprehensive simulation study that demonstrates how concepts from evolutionary game theory \citep{smith1982evolution} can be applied to predict the emergence of cooperation between pairs of state-dependent q-learners that are playing the repeated prisoner's dilemma. Second, it provides evidence that the findings obtained from the simulations are useful in predicting the emergence of cooperation between humans.

The simulation study analyzes the learned strategy profiles between pairs of $\epsilon$-greedy q-learners that have a constant exploration probability and a one-period memory. A strategy profile is considered cooperative if it reinstates cooperation when players are in a state of mutual defection and if it rules out sequences of play in which at least one player is repeatedly exploited; that is, in situations where one player repeatedly cooperates while the other player repeatedly defects.

For q-learning algorithms with a constant learning rate and a constant exploration probability in an interval  between $0.01$ and $0.1$, the simulations demonstrate that the frontier between the parameter space that is dominated by non-cooperative strategies and the parameter space that is inducive to cooperative strategies can be characterized by using the potential function of the stochastic replicator dynamics \citep{foster1990stochastic} of the grim trigger strategy. Potential functions, a concept from mechanical physics, can be used to compute the \textit{kinetic} energies with which the dynamics of the game are attracted towards cooperation and defection. 

The central finding of the simulation study is that cooperative strategies start emerging when the ratio of the kinetic energies attracting the dynamics towards cooperation and defection, simultaneously, approaches a critical value, $\mathcal{C}$, defined by the following equation: 

\begin{equation}\label{eq:crit_int}
	\mathcal{C} = \frac{\delta}{1-\delta}\mathcal{K}(\alpha)\epsilon,
\end{equation}

where $\delta$ denotes the discount factor of the repeated game, and $\alpha$ and $\epsilon$ are the learning rate and the exploration probability of the algorithms, respectively. $\mathcal{K}(\alpha)$ is a correction factor that controls for the effect of the exploration probability on the location of the frontier. $\mathcal{K}(\alpha)$ itself depends on the learning rate and is estimated from the simulated data. The frontier is found to be robust to both optimistic and pessimistic initialization of the q-values. 

The gradient of the share of cooperative strategies as a function of the kinetic energy ratio at the frontier increases with the distance between the game parameters and the hyperplane that characterizes the incentive compatibility ($IC$) constraint for cooperation under grim trigger. When the distance between the game parameters and the hyperplane that characterizes the $IC$ constraint exceeds approximately $50 \%$ of its maximum value, the gradient stabilizes and a sharp increase in cooperative strategies is observed in the vicinity of the critical value. 

To be clear, this article does not provide a theoretically founded characterization of the outcomes obtained between simultaneous q-learners with one-period memory in the repeated prisoner's dilemma. Despite the growing interest in understanding the emergence of cooperation in this type of scenario, no characterization of the corresponding conditions is known to the author. To make progress, this article uses an explorative approach to demonstrate how theoretical concepts, which are plausibly related to q-learning dynamics, can be used to predict the emergence of cooperation. While the motivation for relying on the theory of stochastic replicator dynamics will be discussed, the theoretical characterization of the relationship between stochastic replicator dynamics and simulatenous q-learning with one-period memory is left for future research. 

The ratio of the kinetic energy depends on the discount factor and on \textit{all} of the  payoffs from the prisoner's dilemma. It depends, in particular, on the payoff obtained when the other player deviates from mutual cooperation. This payoff, known as the ``sucker's'' payoff is the main determinant of the risk associated with choosing to cooperate. A body of literature from the neurosciences suggests that reinforcement learning models are useful to understanding human behavior in risky situations \citep{Denrell2007, Shteingart2013, Shteingart2014}. This article elaborates on this nexus by asking how the insights obtained for $\epsilon$-greedy q-learners apply to humans. To this end, the article uses a comprehensive meta dataset about laboratory experiments analyzing human choices in the repeated prisoner's dilemma \citep{dal2018determinants}.

 In this regard, one complicating aspect is the interpretation of the algorithms' exploration probability and its effect on the location of the frontier. The critical value defined in \cref{eq:crit_int} corresponds to the expression $\delta/(1-\delta)$ when the effect of the algorithms' exploration probability cancels out with the correction factor. The approach chosen in this article is to test whether cooperation between humans emerges when the kinetic energy ratio approaches $\delta/(1-\delta)$. A possible interpretation is that this captures the case in which the exploration probability does not affect the location of the frontier. Since exploration leads to occasional random implementation of actions, that is, to noise, ignoring the impact of the exploration probability can be interpreted as assuming a deterministic implementation of actions, which is the scenario studied in the meta data about laboratory experiments with humans. 

The analysis contrasts the predictions derived from the kinetic energy ratio with the predictions derived from $sizeBAD$, which is the conventionally used measure to predict the emergence of cooperation between humans. $sizeBAD$ is connected to the kinetic energy ratio through the potential function of the stochastic replicator dynamics. However, $sizeBAD$ ignores the information about the forces that are simultaneously attracting the system towards cooperation and defection. Therefore, the kinetic energy ratio leads to a more complete ordering over the cooperation incentives induced by the possible payoff structures of the prisoner's dilemma. The results of the meta data analysis reveal that the kinetic energy ratio appears to be a good indicator for the emergence of cooperation between humans. The Pearson correlation coefficients between the cooperation rates observed for humans and those observed for q-learners are consistently above $0.8$.

\subsection{Related Literature}
\label{sec:lit}
\citet{tuyls2005evolutionary} provide an in-depth overview of the connection between multi-agent q-learning and evolutionary game dynamics. The similarity becomes especially clear in the continuous time limit of the q-learning dynamics under Boltzmann exploration, which was first derived in a study by \citet{tuyls2003selection}. \citet{banchioadapt2022} characterize the continuous time limit for $\epsilon$-greedy q-learning algorithms. Both \citet{tuyls2003selection} and \citet{banchioadapt2022} consider the special case of q-learners that learn action rewards without conditioning on the actions taken in the previous period. 

\citet{meylahn2022limiting} characterize the mutual pure strategy best responses in the symmetric two-player, two-action repeated prisoner's dilemma with one-period memory and show that the corresponding best-response dynamics are realized by sample batch q-learning in the infinite batch size limit, which is also considered by \citet{usui2021symmetric}. In contrast to q-learning with infinite batch size limit, in which both agents \textit{alternately} use q-learning while holding fix the strategy of the other player, this article is concerned with characterizing the emergence of cooperation when both agents \textit{simultaneously} learn their strategies.

The simulation study presented in this article relates the observed cooperation rates between state-dependent q-learners with one-period memory to statistics derived from the potential function that describes the stochastic evolutionary replicator dynamics of the grim trigger strategy. The theory of stochastic replicator dynamics, introduced by \citet{foster1990stochastic},  requires robustness to stochastic shocks, incorporating a notion that intuitively appears relevant for understanding the convergence of algorithms that use $\epsilon$-greedy exploration.

The mathematical connection between evolutionary game theory and q-learning and the importance of stochastic shocks in both $\epsilon$-greedy exploration and the theory of stochastic replicator dynamics motivate the approach chosen for the simulation study. While the theory of stochastic replicator dynamics does not incorporate state-dependent action rewards, this article demonstrates that it still allows useful insights with respect to the results that emerge from repeated interactions between state-dependent $\epsilon$-greedy q-learners.

Since the seminal work of \citet{axelrod1981evolution}, a rich body of theoretical literature has studied the emergence of cooperation in the repeated prisoner's dilemma \citep{boyd1987no, boyd1989mistakes}. Different equilibrium concepts, such as evolutionary stable strategies \citep{smith1982evolution} and scholastically stable strategies \citep{foster1990stochastic}, and refinements thereof, have been applied to study this question. \citet{binmore1992evolutionary} apply the concept of evolutionary stability to finite automata games when players incur complexity cost \citep{abreu1988structure} and find that the outcome of such games are cooperative. \citet{volij2002defense}, on the other hand,  finds that continued mutual defection is the only stochastically stable equilibrium with finite automata. The results presented in this article indicate that cooperation emerges under conditions which are more restrictive than the ones implied by stochastic stability.

By analyzing whether the insights obtained from the simulations with q-learning algorithms are useful to predicting the choices humans make when playing the repeated prisoner's dilemma, this article relates to literature from the neurosciences that uses model-free reinforcement learning algorithms, such as q-learning, to model human decision making in risky environments \citep{hertwig2004decisions, Denrell2007, Shteingart2013, Shteingart2014}.\footnote{On a more elementary level, the use of reinforcement learning algorithms to model human learning processes from past experience is supported by neurological studies that support the hypothesis that the mammalian brain itself uses a form of model-free reinforcement learning \citep{montague2004computational, kim2012optogenetic, schultz2013updating}.}

This work is further related to the economic literature that studies the determinants of cooperation in the repeated prisoner's dilemma. While the folk-theorem \citep[see][]{mailath2006repeated} suggests that cooperation is sustainable under fairly mild conditions, experiments such as the early work of \citet{roth1978equilibrium} provide convincing evidence that this game-theoretical frontier is only a poor predictor for the emergence of cooperation in real-world settings.  

This observation has prompted a series of laboratory experiments aimed at better understanding the game parameters that are necessary to induce cooperative behavior.  \citet{blonski2011equilibrium} use the concept of risk dominance to explain the emergence of cooperation. \citet{dal2011evolution} propose $sizeBAD$, a measure that refines the dichotomous concept of risk dominance. The meta dataset used to analyze cooperation between humans is obtained from \citet{dal2018determinants}, which also provides an in-depth overview and meta analysis of the experimental literature that studies the determinants of cooperation.

\cref{sec:ess} discusses how $sizeBAD$ can be derived from the potential function of the stochastic replicator dynamics and how the richer information contained in the kinetic energy ratio is likely to make it a more suitable index to explain the emergence of cooperation. The results presented in this article show how studying q-learning algorithms might provide valuable insights to better understand the conditions under which human actors cooperate. 

\section{Theoretical Background}
\label{sec:ess}
\subsection{The Prisoner's Dilemma}

Throughout this study, the normalized prisoner's dilemma as shown in the matrix on the left-hand side of \cref{fig:payoff_matrix} will be considered. $r$ denotes the normalized reward from cooperation and $-s$ is the normalized ``sucker's'' payoff. The temptation payoff, $t$, is normalized to one, and the stage game Nash equilibrium payoff, $p$, is normalized to zero. The normalization is obtained by subtracting $p$ from the original payoffs and then dividing the resulting value by $t - p$.  This game constitutes a prisoner's dilemma if $1 \geq r \geq 0 \geq -s$. 

\begin{table}[h!]
	\setlength\extrarowheight{6.5pt}
	\begin{minipage}{.475\linewidth}
		\captionsetup{font = footnotesize, margin = {0.75cm,0.5cm}}
		\caption*{Prisoner's Dilemma Payoffs}
		\vspace{-0.5cm}
		\begin{tabular}{c|m{2cm}<{\centering}|m{2cm}<{\centering}|}
			\multicolumn{1}{c}{}  & \multicolumn{1}{c}{\textit{coop}}  & \multicolumn{1}{c}{\textit{defect}} \\ \cline{2-3}
			\textit{coop}               &        $r$                  &    $-s$    \\ \cline{2-3}
			\textit{defect}             &       $1$                  &     $0$         \\ \cline{2-3}
		\end{tabular}
	\end{minipage}%
	\begin{minipage}{.475\linewidth}
		\captionsetup{font = footnotesize, margin = {0.75cm,0.5cm}}
		\caption*{Average Discounted Payoffs GT}
		\vspace{-0.5cm}
		\begin{tabular}{c|m{2cm}<{\centering}|m{2cm}<{\centering}|}
			\multicolumn{1}{c}{}  & \multicolumn{1}{c}{\textit{coop}}  & \multicolumn{1}{c}{\textit{defect}} \\ \cline{2-3}
			\textit{coop}               & $r$                                                  &  $-(1-\delta)s$   \\ \cline{2-3}
			\textit{defect}             &  $(1-\delta)$                                   & $0$      \\ \cline{2-3}
		\end{tabular}
	\end{minipage} 
	\captionsetup{font = small}
	\captionof{figure}{Row Player's Payoffs in a One-Shot Game and Under Grim Trigger}\label{fig:payoff_matrix}
\end{table}

\subsection{Incentive Compatibility Constraint of Grim Trigger}

In the repeated prisoner's dilemma, the grim trigger strategy specifies that if both players \textit{always} cooperated in the past then each player continues cooperating, otherwise, they should defect. The set of all possible histories can therefore be partitioned into two disjoint subsets: the histories for which grim trigger specifies cooperation and the histories for which grim trigger specifies defection.  

Because of its simplicity, and according to the one step deviation principle \citep[see][]{mailath2006repeated}, the subgame perfection of the grim trigger strategy can be verified using the normal form representation depicted in the matrix on the right-hand side of \cref{fig:payoff_matrix}. If the other player defects, then defection is always a best response. By contrast, if the other player cooperates, then cooperation is a best response only if $r \geq (1-\delta) $. Thus, grim trigger is a subgame perfect Nash equilibrium if $r  + \delta -1  \geq  0$. The $IC$ constraint of grim trigger is binding if $r  + \delta -1  =  0$. 

The shortest distance between the tuple $(r,\delta)$ and the hyperplane $r  + \delta -1  =  0$ is given by

\begin{equation}\label{eq:dic}
	d^{ic} = (\delta + r - 1)/\sqrt{2}. 
\end{equation}

\noindent \cref{eq:dic} can be interpreted as a measure of the slackness of the $IC$ constraint. The shortest distance is maximized when $\delta=r=1$; that is, the maximum shortest distance is given by $d^{ic} = 1/\sqrt(2)$. A distance of zero corresponds to a binding $IC$ constraint. Cooperation cannot be supported under grim trigger if the distance is negative, that is, when the $IC$ constraint is violated. Throughout, only tuples $(\delta, r)$ with $d^{ic}>0$ will be considered.

\subsection{Stochastic Replicator Dynamics}

In evolutionary game theory, replicator dynamics characterize the per-period growth rate of different phenotypes in a population, under the assumptions that, in each period, there is a large number of random encounters between individuals with different phenotypes. A phenotype can be understood as a specific evolutionary strategy.  

\textit{Deterministic} replicator dynamics implicitly assume that each encounter results in the same change in the biological fitness, which is captured by a constant payoff. \textit{Stochastic} replicator dynamics acknowledge that the change in biological fitness resulting from an individual encounter is additionally affected by a noise component.

Given the row player's payoff matrix, $A$, the stochastic replicator dynamics can be approximately described by the following Wiener process \citep[the notation is taken directly from][]{foster1990stochastic}:  

\begin{equation}\label{eq:stochastic_dynamics}
	dp_i(t) = p_i(t)\big((Ap(t)\big)_idt - p(t)^{T}Ap(t)dt + \sigma(\mathcal{T}(p)dW(t))_i ).
\end{equation}

\noindent $p_i(t)$ denotes the probability of playing strategy $i$ in period $t$, $W(t)$ is a white noise process, and $\mathcal{T}(p)$ is continuous in $p$. Throughout this study, $A$ will be given by the right-hand side matrix in \cref{fig:payoff_matrix}.

Relying on stochastic replicator dynamics to study the outcomes emerging between q-learners can be motivated by noting that the right-hand side of \cref{eq:stochastic_dynamics} approximately describes the dynamics between q-learners using softmax exploration \citep{tuyls2003selection}. The difference between the replicator dynamics in \cref{eq:stochastic_dynamics} and the dynamics presented in \citet{tuyls2003selection} is in the noise component, which, when using softmax exploration, is related to the differences between the q-values. It should be noted that the tuning parameter of softmax exploration (called the temperature or heat) can always be chosen in such a way that changes in the differences between q-values will only have a negligible effect on the changes of the action selection probabilities, which approximates the case of $\epsilon$-greedy exploration considered in this article. 

While the results presented in \citet{tuyls2003selection} motivate why it appears sensible to consider the stochastic replicator dynamics in \cref{eq:stochastic_dynamics} to study q-learning under $\epsilon$-greedy exploration, the derivations in \citet{tuyls2003selection} do not account for state-dependency, which is a necessary feature to model games with one-period memory. However, it is worthwhile to point out that using the grim trigger payoff matrix in \cref{eq:stochastic_dynamics} introduces a notion of state-dependency because the continuation payoffs of grim trigger assume the play of state-dependent action profiles. Typically, \cref{eq:stochastic_dynamics} is used with stage-game payoff matrices. 

\begin{figure}[h!]
	\centering
	\includegraphics[width=0.65\textwidth]{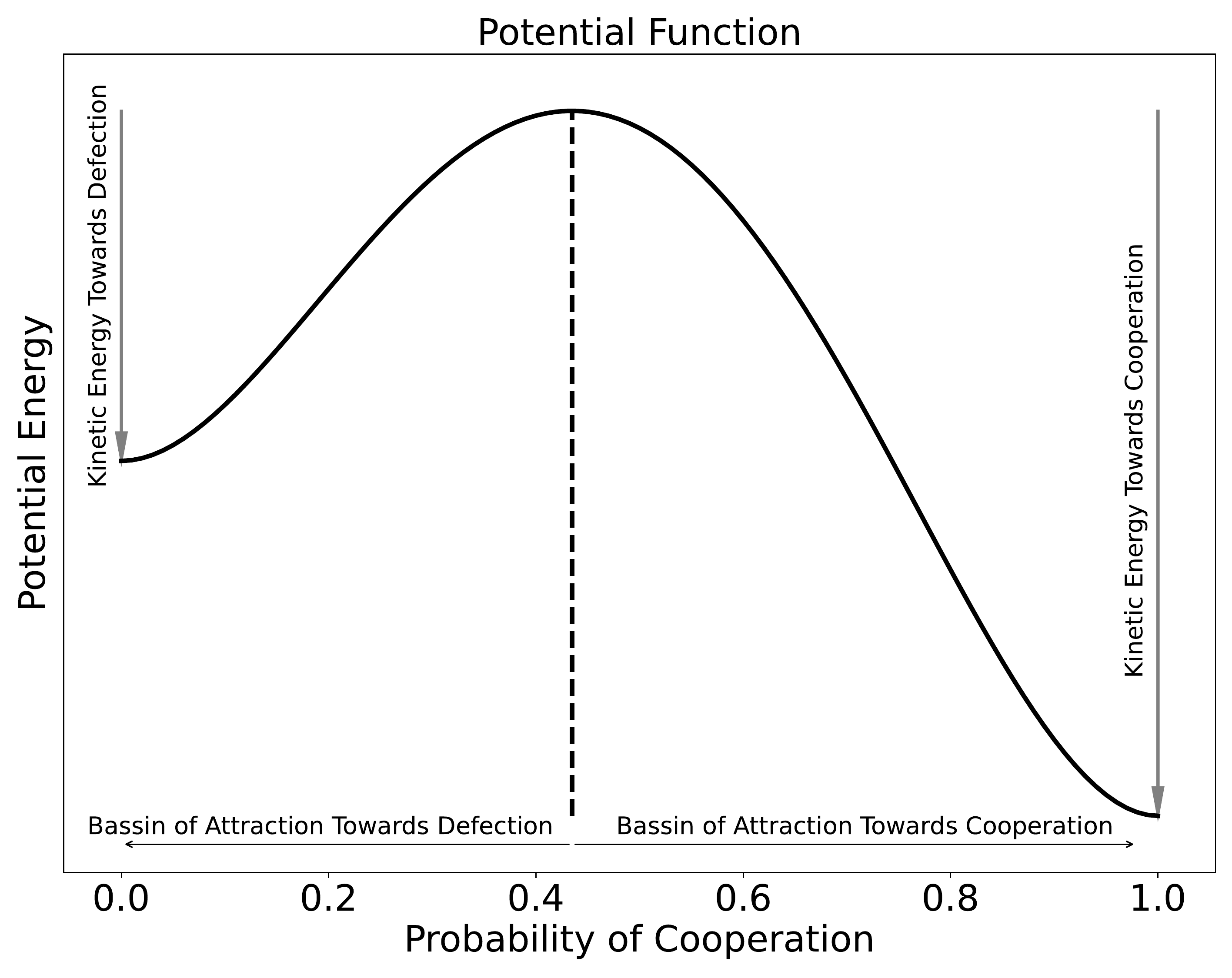}
	\caption{Potential Function under Grim Trigger}\label{fig:kineg_illus}
	\caption*{\footnotesize Note: The figure assumes that $r>(1-\delta)$ and $s > 0$.}
\end{figure}

The potential function, $U(p)$, of the stochastic replicator dynamics is given by

\begin{equation}\label{eq:potfunc}
	U(p) = -\int_0^p x\big((Ax)_i - x^{T}Ax\big)dx. 
\end{equation}

\noindent The potential function is easy to compute in two action games. \cref{app:lem} presents the explicit formulas for the grim trigger strategy. In mechanical physics, potential functions describe the potential energy an object holds because of its position. As the object changes position, potential energy is converted into kinetic energy. An object's kinetic energy resulting from moving from $\underbar{p}$ to $\overline{p}$ is given by $U(\underbar{p}) - U(\overline{p})$. 

\cref{fig:kineg_illus} illustrates the shape of the potential function under the grim trigger strategy when $r>(1-\delta)$ and $s>0$. The domain to the right of $p^{\star}= \argmax_p U(p)$ corresponds to the basin of attraction to cooperation. Intuitively, the domain of attraction to cooperation can be thought of as the set of points from which the system would naturally evolve to full cooperation ($p=1$) by ``moving down the slope'' to the right of $p^{\star}= \argmax_p U(p)$. The basin of attraction to defection corresponds to the domain to the left of $p^{\star}= \argmax_p U(p)$. 

Keeping with the mechanical physics terminology, I denote by $KE^{c} = U(p^{\star}) - U(1)$ the kinetic energy generated by the basin of attraction to cooperation, and by $KE^{d} = U(p^{\star}) - U(0)$ the kinetic energy generated by the basin of attraction to defection. Expanding on the previous intuition, these quantities describe the energy generated by ``moving down the slope.''

\subsection{The Kinetic Log-ratio}

The claim of this article is that the share of cooperative strategies starts increasing when the ratio between the kinetic energies exerted by cooperation and defection approaches the critical value $\mathcal{C} = \frac{\delta}{1-\delta}\mathcal{K}(\alpha)\epsilon$. More precisely, the frontier between cooperation and defection is characterized by the following equality:

\begin{equation}\label{eq:crit}
	\frac{KE^{c}}{KE^{d}}= 	\frac{\delta}{1-\delta}\mathcal{K}(\alpha)\epsilon, 
\end{equation}

\noindent where $\alpha$ denotes the learning rate and $\epsilon$ denotes the exploration probability. For each $\alpha$, $\mathcal{K}(\alpha)$ is a constant that is estimated from the data by searching over a grid of candidate values and selecting the one that minimizes the mean-squared error between the predicted and the observed frontiers.\footnote{The details of the calibration and a tabulation of $\mathcal{K}(\alpha)$ for different values of $\alpha$ are provided in \cref{app:correction_factor}.} It is convenient to use a rearranged log-transformation of \cref{eq:crit}: 

\begin{equation}\label{eq:klr}
	\underbrace{log\big((1-\delta)KE^{c}\big) - log\big(\delta KE^{d} \big)}_{\mathcal{KLR}} = log(\mathcal{K}(\alpha)\epsilon).
\end{equation}

The left-hand side of \cref{eq:klr}, henceforth referred to as the \textit{kinetic log-ratio} ($\mathcal{KLR}$), only depends on the parameters of the game, while the right-hand side only depends on the algorithms' tuning parameters.

\subsection{Properties of the Potential Function Under Grim Trigger}

\cref{lem:properties} summarizes the relevant properties of the potential function and the statistics derived from it under the grim trigger strategy.\footnote{The corresponding proofs are presented in \cref{app:lem}.} Additionally, it establishes the connection between the potential function and $sizeBAD$, the measure used in the experimental economic literature studying the emergence of cooperation between humans.

Like the potential function considered in this article, the definition of $sizeBAD$ \citep{dal2011evolution} relies on the grim trigger strategy. More precisely, using the payoffs of the normalized prisoner's dilemma, $sizeBAD$ is given by the probability $p$ which solves the following equation: 

\begin{equation}\label{eq:sbm}
	p \times \frac{r}{1-\delta} + (1-p) \times (-s) = p \times 1 + (1-p) \times 0.
\end{equation}

The left-hand side of \cref{eq:sbm} denotes the expected value from choosing to cooperate and subsequently playing grim trigger if the opponent randomizes between cooperation (under grim trigger) with probability $p$ and perpetual defection with probability $1-p$. The right-hand side denotes the analogously defined expected value of choosing continued defection.        

Property (i) of \cref{lem:properties} establishes the formal connection between \cref{eq:potfunc} and $sizeBAD$. Property (ii) describes the scenario depicted in \cref{fig:kineg_illus}. In this case, $sizeBAD$ simply corresponds to the domain of attraction to defection. According to $sizeBAD$, the incentive to cooperate increases when the domain of attraction to defection decreases.

Property (iii) states that when the $IC$ constraint of grim trigger is violated, the potential function is strongly increasing over all $p\in[0,1]$, which implies that the entire domain is attracted to defection. In this case, $p^{\star}$ lies outside the unit interval and the basin of attraction to cooperation is the empty set. Property (iv) describes the opposite case where the potential function is decreasing over the entire domain, which occurs when the sucker's payoff is zero and the $IC$ is strictly fulfilled. 

\begin{lemma} \label{lem:properties}
	\begin{itemize}
		\item[] 
		\item[(i)]    $p^{\star}$, which solves $\partial (U/p)/\partial p = 0$, corresponds to $sizeBAD$
		\item[(ii)]  If $r > (1-\delta)$ and $s>0$,  then $p^{\star} = \argmax_p U(p) \in (0,1)$, and  $U(p)$ has two minima at $p = 0$ and $p = 1$ for $p \in [0,1]$
		\item[(iii)]   If $r < (1-\delta)$ and $s>0$, then $\partial U(p)/ \partial p > 0$ for $p \in [0,1]$ and $p^{\star} \notin [0,1]$
		\item[(iv)]  If $r > (1-\delta)$ and $s=0$, then $\partial U(p)/ \partial p < 0$ for $p \in [0,1]$ and $p^{\star} =0$
		\item[(v)]  If $r > (1-\delta)$ and $s \geq 0$, then $\partial \mathcal{KLR}/\partial r > 0$ and $\partial \mathcal{KLR}/\partial s < 0$ 
	\end{itemize}
\end{lemma}

Property (v) shows that the $\mathcal{KLR}$ has desirable properties as an index to measure the propensity to cooperate; for example, the kinetic log-ratio is strictly increasing in the reward from cooperation, $r$, for all suckers' payoffs, $s \geq 0$. This is not the case for $sizeBAD$, which does not indicate increasing cooperation incentives in $r$ when $s=0$. As a result, by ignoring the information contained in the kinetic energy, $sizeBAD$ induces a less complete ordering over the different possible payoff structures of the prisoner's dilemma.

\section{Cooperation Between Q-Learners}
\label{sec:sim}
\subsection{Multi-Agent Q-Learning in the Prisoner's Dilemma}\label{subsec:simd}

This subsection focuses on the aspects relevant to the simulation study. A more in-depth review of q-learning \citep{watkins1989learning} is provided in \cref{app:ql}. The simulations are performed using $\epsilon$-greedy q-learning algorithms with a constant exploration probability. The updating rule applied to the q-values is given by

\begin{equation}\label{eq:qupdtmt}
	Q_{t+1}(s,a) = Q_{t}(s,a) + \alpha \big(u(a,s) + \delta \max_{a \in A} Q_t(s',a) - Q_{t}(s,a) \big),
\end{equation}

\noindent where $s$ denotes the state in period $t$ and $s'$ is the state reached in $t+1$. $a$ denotes the action chosen by the algorithms. $\epsilon$ denotes the exploration probability and $\alpha$ is the learning rate. For both tuning parameters, $\alpha$ and $\epsilon$, the grid $\{0.01,0.02,\cdots,0.1\}$ is considered. Throughout the paper, only the symmetric setting is studied; that is, the tuning parameters of the two algorithms are always identical. 

The q-learners have one-period memory. The state in period $t$, $s \in S$, is determined by the players' most recent actions in $t-1$, $a \in A = \{coop, defect\}$. Hence, the state space, $S$, is determined by the four possible combinations of actions chosen by the two algorithms. The action rewards, $u$, are given by the prisoner's dilemma payoffs shown in the matrix on the left-hand side of \cref{fig:payoff_matrix}. 

\subsection{Simulation Details}

For each tuple $(\alpha, \epsilon)$, the cooperation rates are evaluated over a grid \textbf{\textdelta} $\times$ $ \mathbf{r}  \times \mathbf{s}$. For $\delta$ and $r$, the grid points are given by $\{0.525,0.575,\cdots,0.975\}$. The different combinations of $\delta$ and $r$ generate $25$ distinct distances ($d^{ic}$ defined in \cref{eq:dic}) to the hyperplane which characterizes the binding $IC$ constraint. For each tuple $(\delta,r)$, fifteen values of $s$ generating kinetic log-ratios in the interval $[-5,10]$ are sampled. A stratified sampling procedure is used to ensure that the fifteen intervals, defined by the adjacent integers in $[-5, 10]$, each contain one sampled value. The range of $\mathcal{KLR}$ values considered is chosen in such a way that, for each $(\alpha,\epsilon)$ tuple considered, the re-centered values, $\mathcal{KLR} - \mathcal{K}(\alpha)\epsilon$, span the entire interval $[-5,5]$. In total, $150{,}000$ parameter combinations are considered. For each combination of parameters, $100$ games are simulated. The q-values are initialized using the infinite discounted sum of the cooperation reward, and each game lasts for one million periods.\footnote{The chosen initialization procedure corresponds to an optimistic approach. In \cref{app:init_pes}, the case of pessimistic initialization, using the stage game Nash equilibrium payoff of zero, is considered. Additionally, games lasting five million periods are considered for selected parameter combinations. Both robustness analyses deliver similar results to the ones presented in the main analysis. With pessimistic initialization, the observed cooperation rates are smaller, as is the gradient at the frontier. Nevertheless, the frontier remains accurate in predicting the emergence of cooperative strategies.}

\subsection{Cooperative Strategies}

After a game is completed, the last q-matrices observed for both players are used to compute the transition matrix between the four possible states when both players follow their learned strategies. The definition used to compute the share of cooperative strategy profiles for each parameter combination is as follows:  

\begin{definition}[Cooperative Strategy Profile]\label{def:coop}
	The learned strategies are cooperative if mutual defection is not an absorbing state. Additionally, starting from any state, following the learned strategies cannot result in infinite action sequences in which either player suffers the sucker's payoff more than fifty percent of the time. 
\end{definition}

\cref{def:coop} requires that the learned strategies reinstate cooperation after mutual defection. Additionally, the definition rules out situations with one player is continuously defecting while the other is cooperating. \cref{def:coop} can be verified by simulating action sequences based on the transition probability matrix.\footnote{It is noteworthy that \cref{def:coop} rules out scenarios in which both players follow a grim trigger strategy. Another clear regularity, similar to the frontier described by \cref{eq:crit}, could not be established when including unforgiving strategies like grim trigger.} \cref{def:coop} emphasizes robustness to a one-time unilateral deviation from cooperation. This is in contrast to the grim trigger strategy, which can be viewed as \textit{enforcing} behavior through the threat of imposing a maximally harsh punishment. With trigger, cooperation irrevocably collapses in the event of a one-time unilateral deviation.

\begin{figure}[t!]
	\centering
	\begin{subfigure}[b]{0.46\textwidth}
		\centering
		\includegraphics[width=\textwidth]{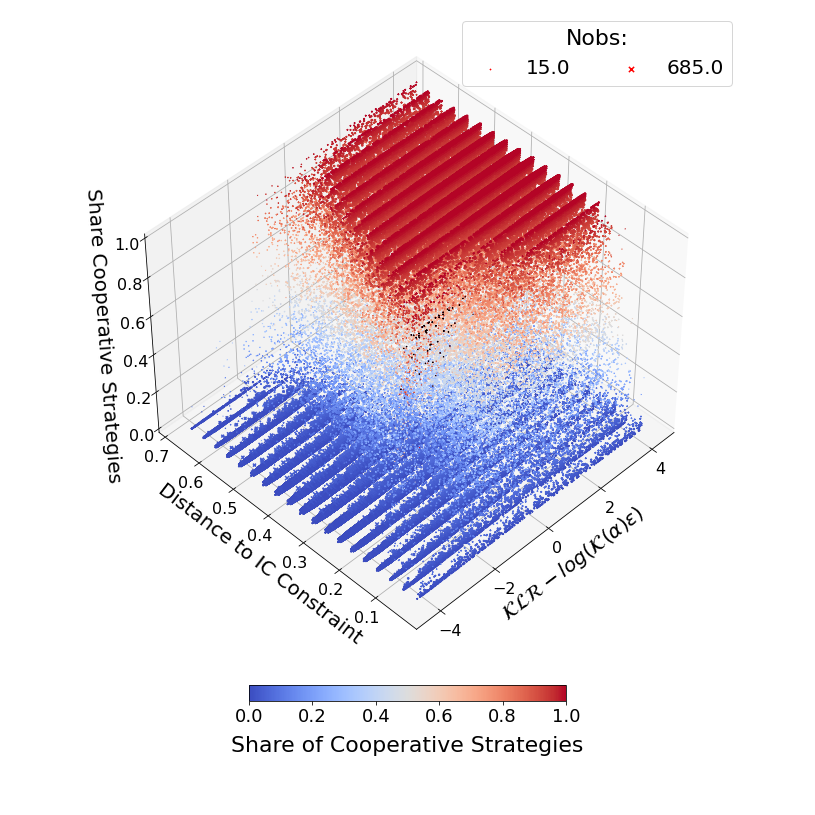}
		\caption{3D Scatter -- All parameters}
		\label{sfig:sim_all}
	\end{subfigure}
	\hfill
	\begin{subfigure}[b]{0.46\textwidth}
		\centering
		\includegraphics[width=\textwidth]{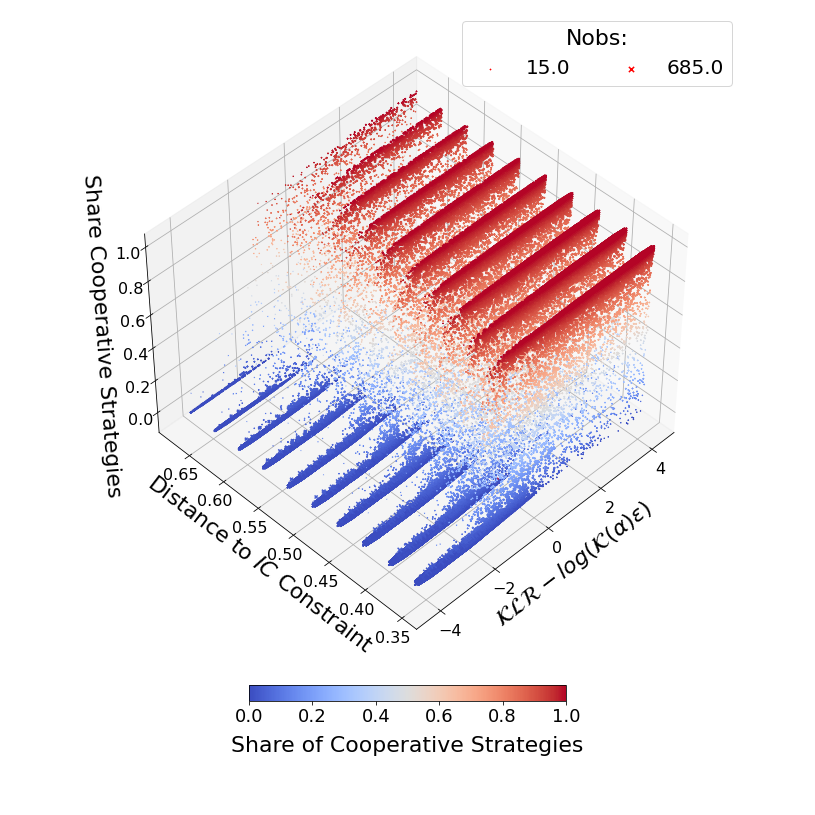}
		\caption{3D Scatter -- Slack $IC$}
		\label{sfig:sim_slack_ic}
	\end{subfigure}
	\par\medskip
	\par\medskip
	\begin{subfigure}[b]{0.46\textwidth}
		\centering
		\includegraphics[width=\textwidth]{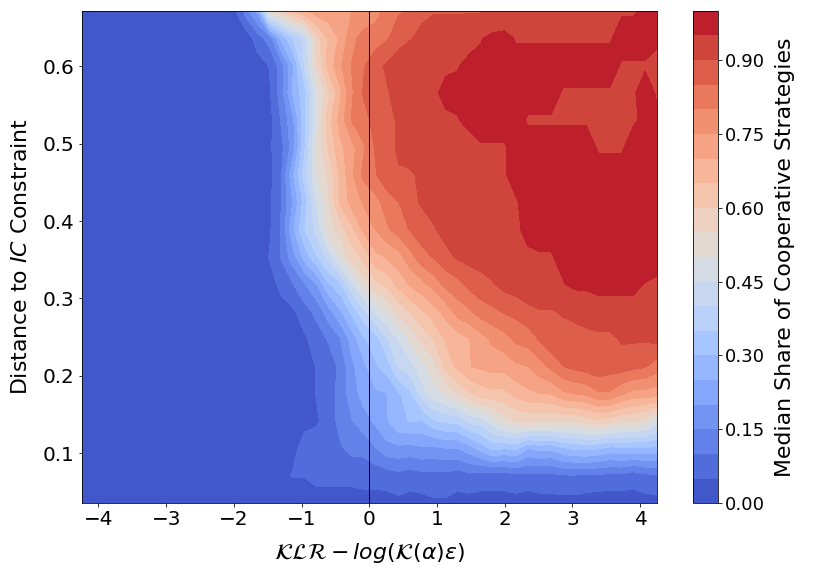}
		\caption{Contour Plot -- All parameters}
		\label{sfig:hm_all}
	\end{subfigure}
	\hfill
	\begin{subfigure}[b]{0.46\textwidth}
		\centering
		\includegraphics[width=\textwidth]{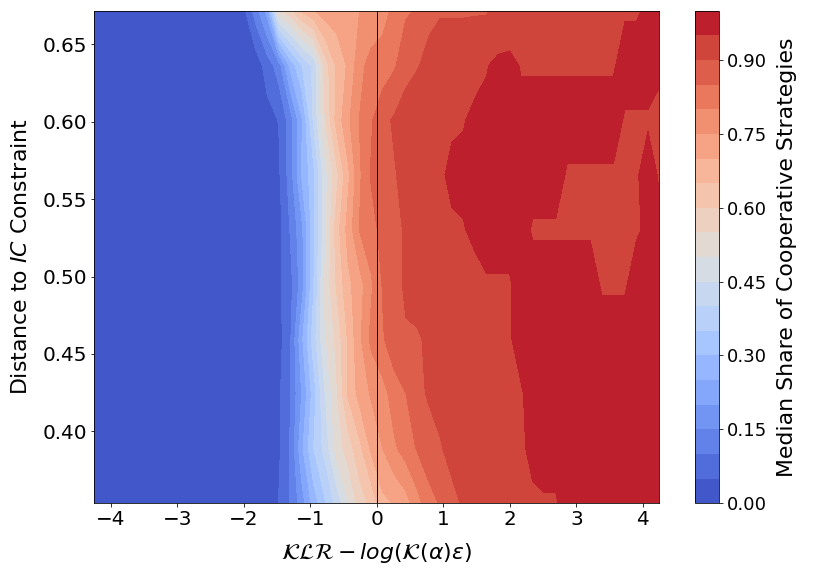}
		\caption{Contour Plot -- Slack $IC$}
		\label{sfig:hm_slack_ic}
	\end{subfigure}
	\caption{Simulation Results}
	\label{fig:sim_res}
	\caption*{\footnotesize Note: The size of each marker in \cref{sfig:sim_all} and \cref{sfig:sim_slack_ic} is calculated by first normalizing all of the variables to lie in the unit interval and then counting the number of neighboring observations in an open ball with radius $0.05$. Outliers with cooperation rates close to one near the binding $IC$ constraint are colored in gray (\cref{sfig:sim_all}). The contour plots in \cref{sfig:hm_all} and \cref{sfig:hm_slack_ic} show the isoquants for the median share of cooperative strategies computed in an open ball of radius $0.05$ around grid points of the normalized explanatory variables. The grid is based on $50$ equally spaced $\mathcal{KLR} - log(\mathcal{K}(\alpha)\epsilon)$ values for each distance to the binding $IC$ constraint.}
\end{figure}

\subsection{Results}\label{subsec:results}

\cref{sfig:sim_all} relates the share of cooperative strategies to the $\mathcal{KLR}$ and to the distance between the game parameters and the binding $IC$ constraint. The $\mathcal{KLR}$ is re-centered around the correction factor $\mathcal{K} (\alpha) \epsilon$. The size of each scatter point indicates the number of observations in an open ball around that point. \cref{sfig:sim_slack_ic} focuses on parameter constellations with a distance to the binding $IC$ constraint that exceeds $50\%$ of its possible maximum value. \cref{sfig:hm_all} and \cref{sfig:hm_slack_ic} show the corresponding contour plots for the \textit{median} of the share of cooperative strategies observed in an open ball centered around selected grid points in the space spanned by the $\mathcal{KLR}$ and $d^{ic}$. The median is calculated across all $(\alpha,\epsilon)$ combinations. \cref{eq:klr}, $\mathcal{KLR} - \mathcal{K} (\alpha) \epsilon = 0$, reliably predicts the emergence of cooperative strategy profiles. 

The gradient of the share of cooperative strategies as a function of the $\mathcal{KLR}$ increases as the distance to the binding $IC$ constraint becomes larger. When this distance exceeds $50\%$ of its maximum possible value, the gradient stabilizes and the critical value characterizes a sharp frontier. In general, the observed variation of the share of cooperative strategies at the frontier reduces with the distance to the binding $IC$ constraint. When the game parameters approach the $IC$ constraint, outliers (colored in gray in \cref{sfig:sim_all}) with high cooperation rates \textit{before} the $\mathcal{KLR}$ frontier can be observed. A closer analysis reveals that these outliers at the $IC$ constraint arise when the learning rate, $\alpha$, is low. As is shown in \cref{app:het}, except for this anomaly observed close to the binding $IC$ constraint, the tuning parameters do not appear to have a systematic impact on the gradient observed at the frontier.

\begin{figure}[h!]
	\centering
	\begin{subfigure}[b]{0.485\textwidth}
		\centering
		\includegraphics[width=\textwidth]{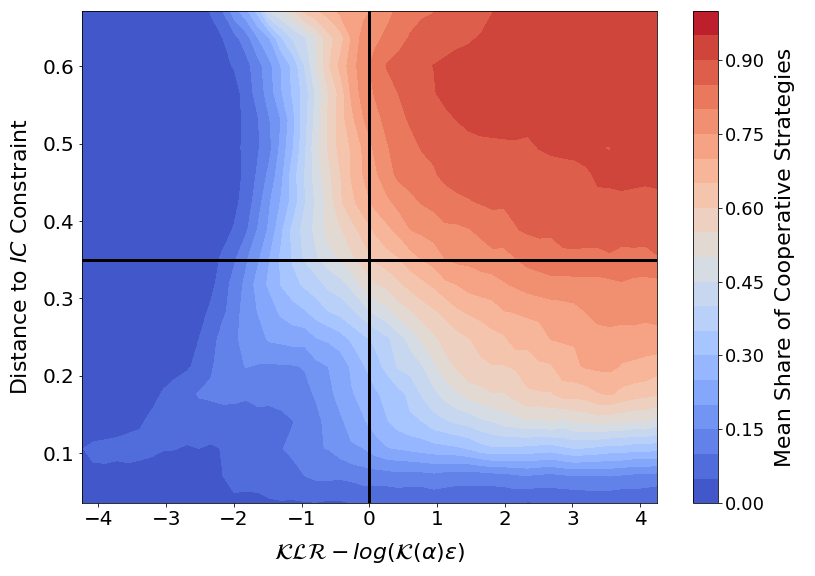}
		\caption{All Cooperative Strategies}\label{sfig:mean_all}
	\end{subfigure}
	\hfill
	\begin{subfigure}[b]{0.485\textwidth}
		\centering
		\includegraphics[width=\textwidth]{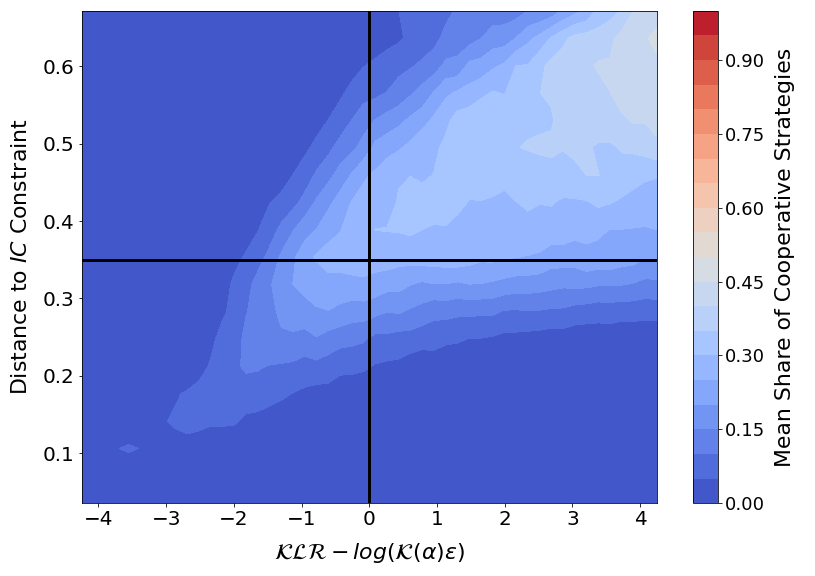}
		\caption{Mutual $ALLC$}\label{sfig:mean_allc}
	\end{subfigure}
	\par\medskip
	\par\medskip
	\begin{subfigure}[b]{0.485\textwidth}
		\centering
		\includegraphics[width=\textwidth]{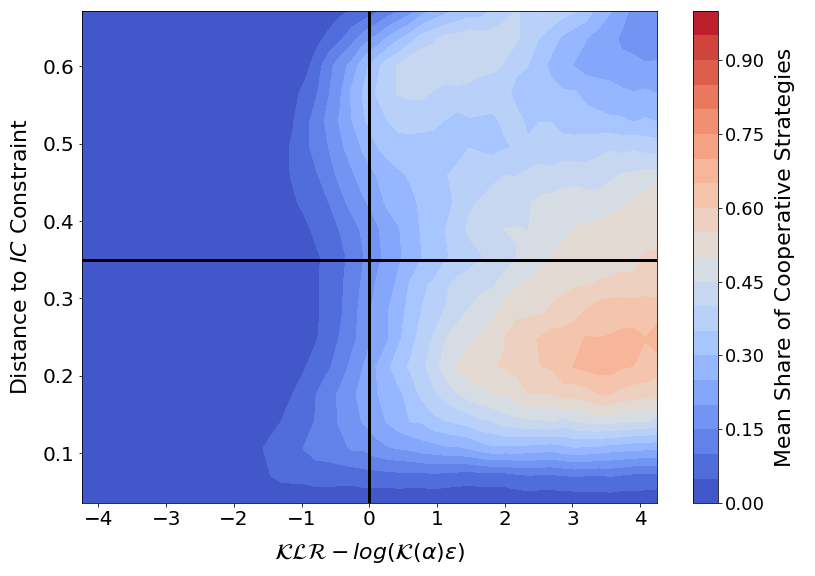}
		\caption{Mutual $WSLS$}\label{sfig:mean_wsls}
	\end{subfigure}
	\hfill
	\begin{subfigure}[b]{0.485\textwidth}
		\centering
		\includegraphics[width=\textwidth]{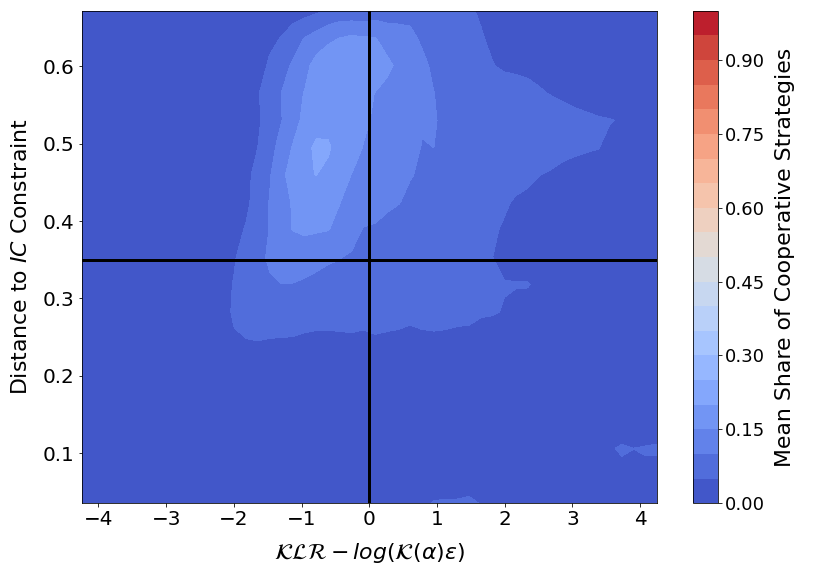}
		\caption{Mutual $OSC$}\label{sfig:mean_osc}
	\end{subfigure}
	\caption{Simulation Results -- Decomposing Cooperative Strategies}
\end{figure}

The \textit{mean} of the share of all cooperative strategy profiles according to \cref{def:coop} is shown in \cref{sfig:mean_all}.\footnote{The mean is calculated in the same way as the median in \cref{sfig:hm_all}.} When using the mean, the anomalies observed when the distance to the binding $IC$ constraint is small become apparent. The cooperative strategy profiles mainly consist of three distinct strategy profiles: (i) mutual $ALLC$ (always cooperate), which specifies that both players always cooperate; (ii) mutual $WSLS$ (win-stay loose-shift), which specifies that both players cooperate on the equilibrium path and that a deviation by either player is punished by mutual play of defection for one period; (iii) mutual $OSC$ (oscillate), which specifies that both players alternate between mutual defection and mutual cooperation. The mean of the share of the respective strategy profiles is shown in \cref{sfig:mean_allc}, \cref{sfig:mean_wsls}, and \cref{sfig:mean_osc}, respectively.\footnote{\cref{app:mean_het} provides a decomposition of the non-cooperative strategy profiles.} 

It is noteworthy that the frontier accurately predicts the emergence of mutual $WSLS$ across almost all distances to the binding $IC$ constraint. When the distance to the $IC$ constraint exceeds $50\%$ of its maximum possible value, $WSLS$ tends to be displaced by $ALLC$ as the $\mathcal{KLR}$ increases. Note that the anomalies close to the hyperplane characterizing the binding $IC$ constraint are not explained by either of the three strategy profiles which dominate in the remainder of the parameter space and which are characterized by an identical choice of strategy by both algorithms.

\vspace{2cm}

\begin{figure}[h!]
	\centering
	\begin{subfigure}[b]{0.485\textwidth}
		\centering
		\includegraphics[width=\textwidth]{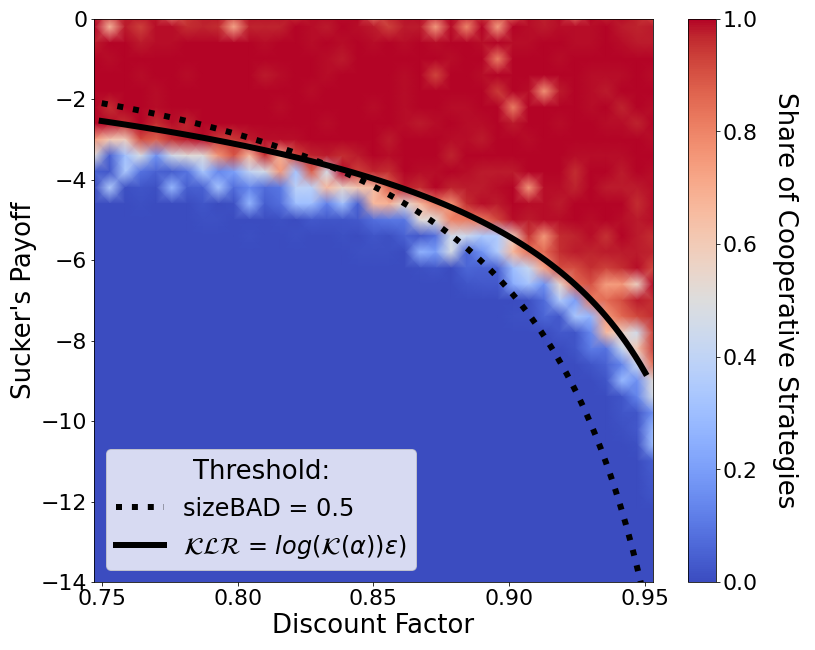}
		\caption{$r=0.775$}\label{sfig:hm_payoff_space1}
	\end{subfigure}
	\hfill
	\begin{subfigure}[b]{0.485\textwidth}
		\centering
		\includegraphics[width=\textwidth]{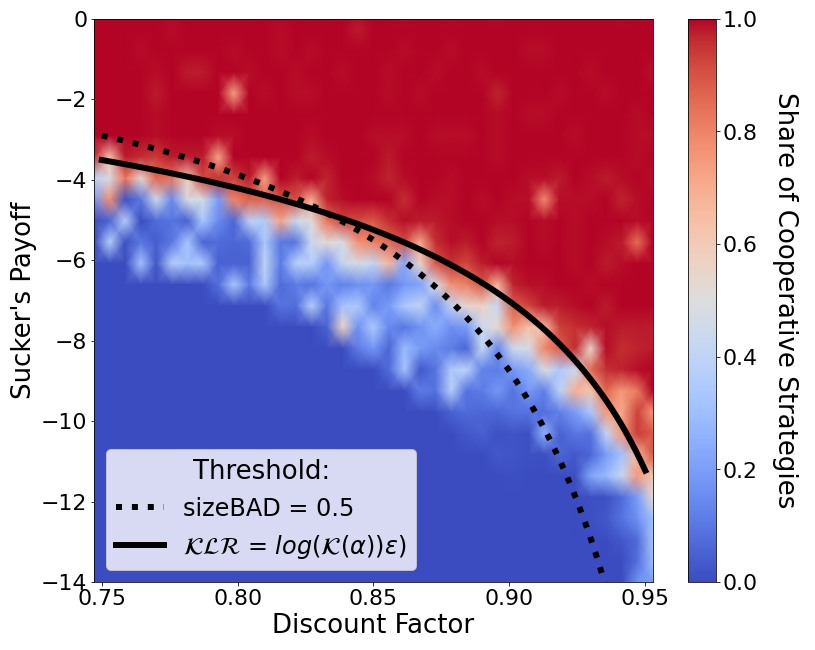}
		\caption{$r=0.975$}\label{sfig:hm_payoff_space2}
	\end{subfigure}
	\caption{Frontier in the $(\delta,s)$ parameter space for $\alpha=0.01$ and $\epsilon=0.01$}\label{fig:hm_payoff_space}
\end{figure}

\cref{sfig:hm_payoff_space1} and \cref{sfig:hm_payoff_space2} show the results obtained from a different set of simulations to illustrate how the frontier maps into the parameter space of the repeated game. The $x$-axes show the discount factor and the $y$-axes denote the sucker's payoff. The simulations rely on $37 \times 37$ grid, which is equally spaced in both dimensions. Hundred games were simulated for each grid point. The cooperation reward is held fixed at a value of $r=0.775$ in \cref{sfig:hm_payoff_space1} and a value of $r=0.975$ in \cref{sfig:hm_payoff_space2}. The learning rate and the exploration probability are both equal to $0.01$. The solid lines denote the $\mathcal{KLR}$ frontiers, while the dotted lines denote the frontiers obtained for $sizeBAD=0.5$.  \cref{fig:hm_payoff_space} illustrates how the $\mathcal{KLR}$ isoquants are superior to the $sizeBAD$ isoquants in capturing the frontier between cooperation and non-cooperation between q-learners.

\section{Cooperation Between Human Players}
\label{sec:labdata}
\subsection{The Data}

This analysis relies on the study by \citet{dal2018determinants} which compiles data from various laboratory experiments that record individuals' actions in repeated prisoner's dilemmas with deterministic choice implementation and perfect monitoring. Only games for which cooperation can be sustained as a subgame perfect Nash equilibrium under grimm trigger will be considered. In total, these experiments contain $24$ different tuples $(\delta, r, s)$, henceforth called treatments.\footnote{In laboratory experiments, the effect of $\delta$ is emulated through random termination of the game. For example, when the discount factor is $0.75$, the game is terminated with probability $0.25$ after each round. Summary statistics about the treatments of each study comprised in the meta dataset are provided in \cref{app:sum_stat}.} 

Typically, each individual participating in a study plays several repeated prisoner's dilemmas in a given treatment. Because of learning effects, it is preferable to study the choice behavior after each individual has played several games. Following the approach of \citet{dal2018determinants}, only individuals' choices observed in the first round of the seventh game will be considered for the main exposition. This allows us to retain all treatments contained in the data. 

\subsection{Comparing the Kinetic Log-ratio and sizeBAD}

The experiments contained in the meta data only refer to prisoner's dilemmas with \textit{deterministic} implementation of actions. This raises the question how to deal with the noise term in \cref{eq:klr}. The approach chosen in this article is to assume that, in the absence of noise, the frontier between cooperation and defection is only determined by the $\mathcal{KLR}$, that is, the correction term on the right-hand-side of \cref{eq:klr} is set to zero.

Technically, the correction factor on the right-hand-side of \cref{eq:klr} is minus infinity when $\epsilon = 0$, which implies that the solution to \cref{eq:klr} is not defined in this case. However, note that the theory behind the computation of the $\mathcal{KLR}$ assumes a vanishingly small noise component. This suggests that only considering the $\mathcal{KLR}$ in \cref{eq:klr} might be the appropriate approach in the absence of noise. Admittedly, this raises the question how to interpret a negative correction factor, which means that the frontier is more lenient than what is implied under vanishing noise. In this regard, it is noteworthy that for the $(\alpha,\epsilon)$ tuples considered in the simulations, the correction factor is predominantly positive and comparatively small in magnitude when negative.\footnote{The minimum correction factor is $-1.16$, while the maximum correction factor is $4.96$} The case of negative correction factors therefore warrants further research and is left for future studies. 

The propensity for cooperative behavior is \textit{increasing} in the $\mathcal{KLR}$ and decreasing in $sizeBAD$. Therefore, to allow for an easier comparison between both measures, it is convenient to define $sizeGOOD = 1- sizeBAD$. For $sizeGOOD$, cooperation is predicted to emerge for values larger than $0.5$, as this signifies that the domain of attraction to cooperation is larger than the domain of attraction to defection.

\begin{proposition}\label{prop:klr_sg}
	If $\delta>0.5$, $r>(1-\delta)$ and $s>0$, the set of tuples $(\delta, r, s)$ for which $\mathcal{KLR} \geq 0$ is a strict subset of the set of tuples $(\delta, r, s)$ for which $sizeGOOD\geq0.5$.
\end{proposition}

\cref{prop:klr_sg} states that the set of parameters for which cooperative behavior is predicted by setting $\mathcal{KLR}$ equal to zero is a strict subset of the set of parameters for which cooperative behavior is predicted by setting $sizeGOOD$ equal to $0.5$.\footnote{The proof is given in \cref{app:lem}.} Thus, $\mathcal{KLR} = 0$ imposes stricter conditions on the emergence of cooperative behavior than $sizeGOOD = 0.5$ when the $IC$ of grim trigger is satisfied and when $\delta > 0.5$.

\subsection{Results}

\begin{figure}[h!]
	\centering
	\begin{subfigure}[b]{0.45\textwidth}
		\centering
		\includegraphics[width=\textwidth]{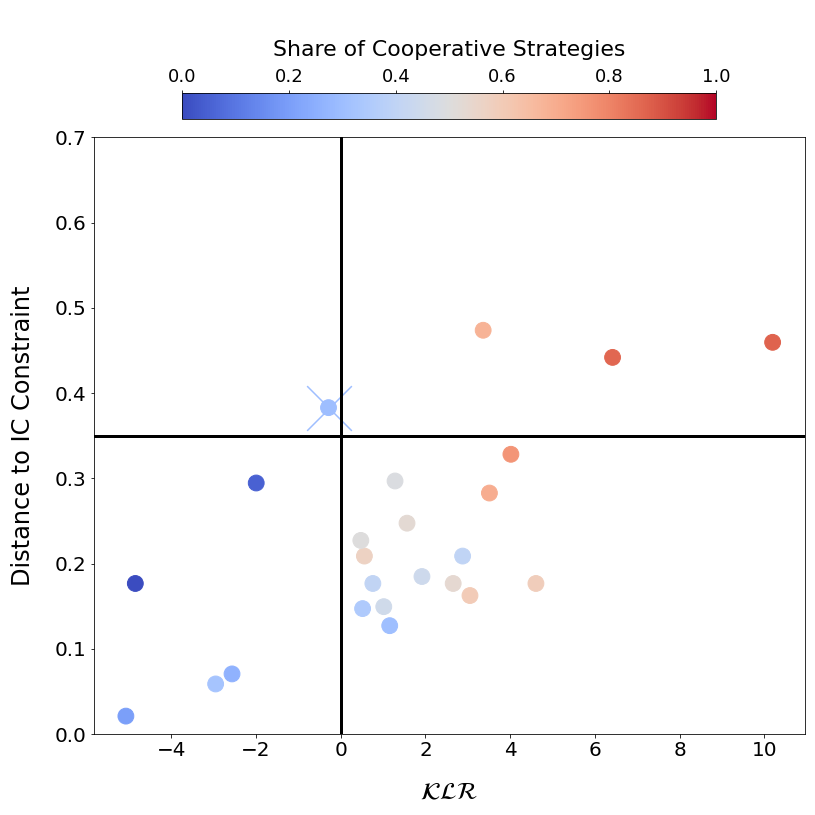}
		\caption{Kinetic log-ratio}
		\label{sfig:human_klr}
	\end{subfigure}
	\hfill
	\begin{subfigure}[b]{0.45\textwidth}
		\centering
		\includegraphics[width=\textwidth]{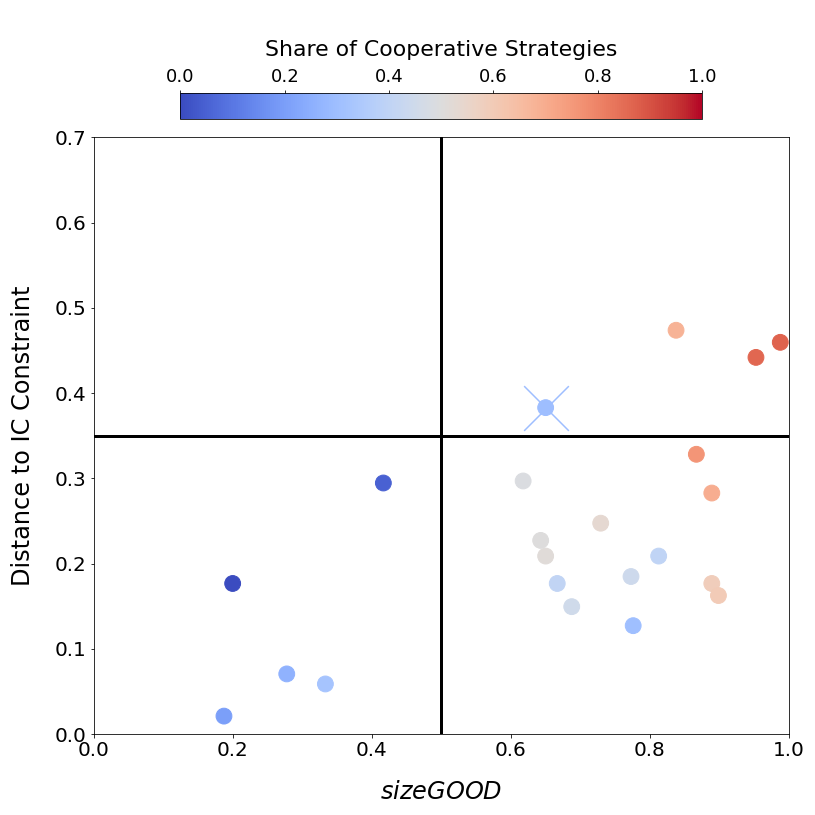}
		\caption{$sizeBAD$}
		\label{sfig:human_sg}
	\end{subfigure}
	\caption{Relation between Cooperation Rates and Two Indices}
	\label{fig:human_res}
\end{figure}

The scatter plot in \cref{sfig:human_klr} relates the observed cooperation rates of each treatment to the $\mathcal{KLR}$ and to the distance to the binding $IC$ constraint. The cooperation rate is defined as the share of individuals choosing to cooperate in the first round of their seventh game in a specific treatment.\footnote{In the experimental literature, focusing on the first round is justified based on the fact that this guarantees the same number of observations by game, irrespective of the continuation probability. Additionally, the literature points out that  first round behavior is not ``contaminated'' by prior actions in the same game, which could lead to dependencies which are difficult to account for. A robustness analysis relying on all rounds to calculate cooperation rates, which is presented in \cref{app:all_rounds}, reveals that the choice of how many rounds to consider is inconsequential for the results.} \cref{sfig:human_sg} repeats the same analysis but uses $sizeGOOD$ on the $x$-axis. The blue shaded dots denote cooperation rates that are below $50\%$ and red shaded dots denote cooperation rates above $50\%$. The observation marked with a large cross designates the only treatment for which we have $\mathcal{KLR}<0$ and $sizeGOOD>0.5$.  In other words, the $\mathcal{KLR}$ predicts conditions that are unfavorable to cooperation and $sizeGOOD$ predicts conditions that are favorable to cooperation. 

The vertical lines denote the respective frontiers for the $\mathcal{KLR}$ and $sizeGOOD$. The horizontal lines denote the distance to the binding  $IC$ constraint for which the $\mathcal{KLR}$ frontier becomes sharp in the simulations with q-learners. Under the assumptions that the frontiers obtained with the simulations hold valuable insights for humans, high cooperation rates should be observed \textit{everywhere} in the upper-right quadrant and low cooperation rates \textit{everywhere} in the upper- and lower-left quadrants. For the lower-right quadrant, we would expect cooperation rates to \textit{gradually increase} as we move away from the $\mathcal{KLR}$ frontier and from the binding $IC$ constraint. 

\begin{figure}[h!]
	\centering
	\begin{subfigure}[b]{0.45\textwidth}
		\centering
		\includegraphics[width=\textwidth]{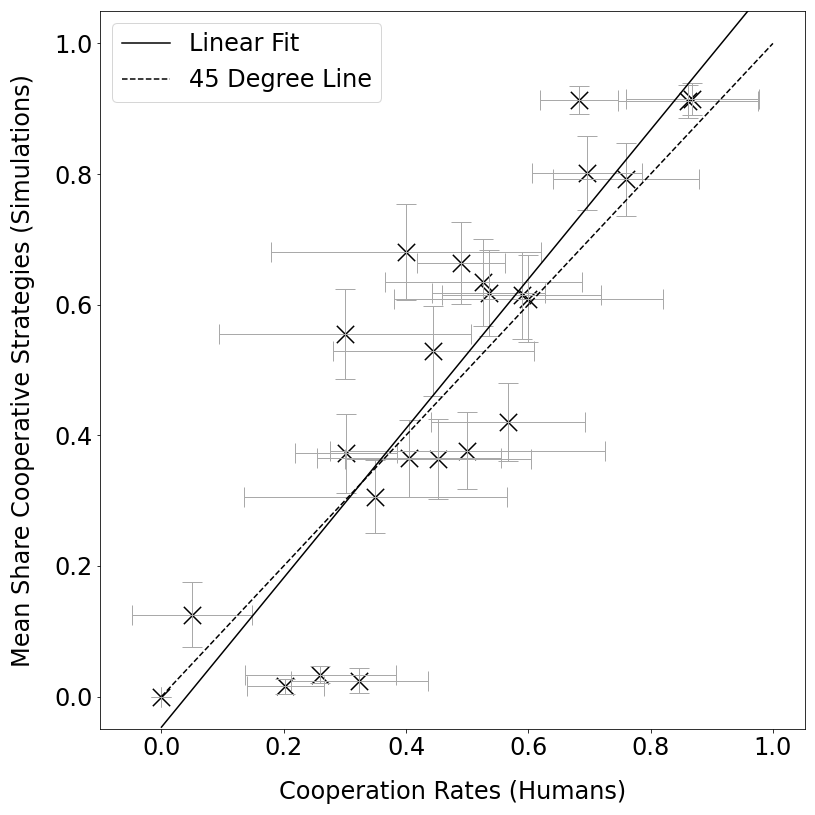}
		\caption{Relation Mean Cooperation Rates  (Game 7)}\label{sfig:corrplot}
	\end{subfigure}
	\hfill
	\begin{subfigure}[b]{0.45\textwidth}
		\centering
		\includegraphics[width=\textwidth]{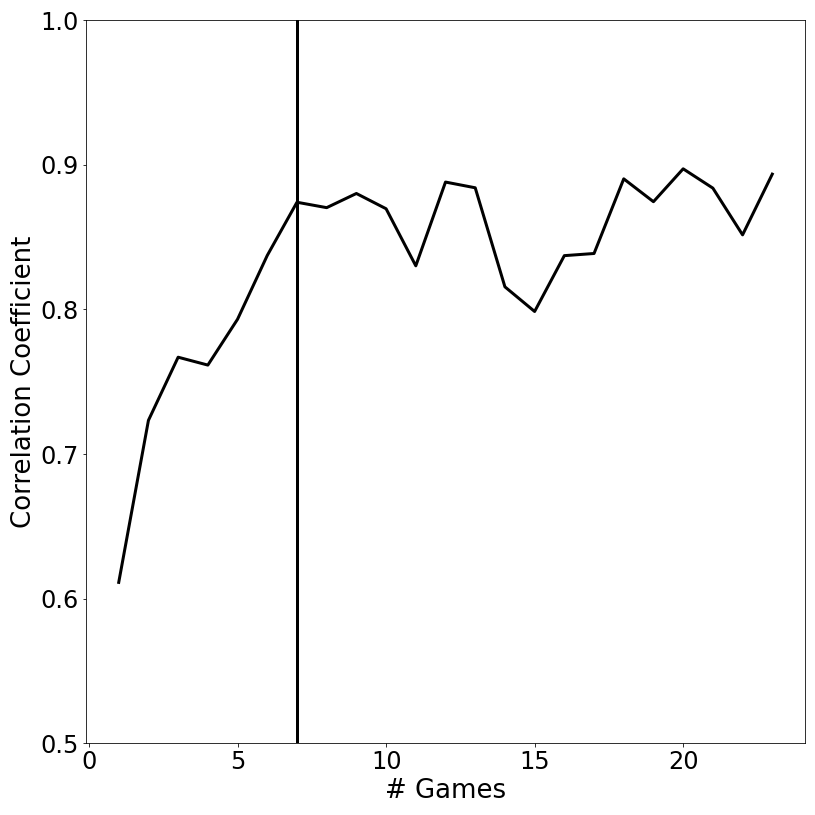}
		\caption{Evolution of Correlation Coefficient}\label{sfig:correvo}
	\end{subfigure}
	\caption{Correlation Between Cooperation Rates Observed for Algorithms and Q-learners}
	\label{fig:corrplot}
\end{figure}

In \cref{sfig:human_klr}, all treatments in the upper-right-hand quadrant have high cooperation rates. This is not the case in \cref{sfig:human_sg}, where we observe a cooperation rate below $50\%$ for the treatment with conflicting predictions according to the $\mathcal{KLR}$ and $sizeGOOD$. While one treatment is not sufficient to reach a definitive conclusion, it is noteworthy that the only data point allowing a systematic test of both measures corroborates the frontier implied by the $\mathcal{KLR}$. It is also noteworthy that the treatments with low cooperation rates in the lower-right-hand quadrant of \cref{sfig:human_klr} tend to be close to the $\mathcal{KLR}$ frontier. By contrast, in \cref{sfig:human_sg}, low cooperation rates in the lower-right-hand quadrant are observed even for treatments with a $sizeGOOD$ measure that is around $0.8$, which is close to its maximum value of one. 

\cref{sfig:corrplot} shows the relationship between the mean share of cooperative strategies observed for q-learning algorithms (y-axis) and the human cooperation rates for each treatment (x-axis). The error bars show the respective $95\%$ confidence intervals. The mean share of cooperative strategies for the algorithms is computed using the $100$ observations which are closest to the treatment of each laboratory experiment when calculating the euclidean distance in the $(d^{ic}, \mathcal{KLR} - log(\mathcal{K}(\alpha)\epsilon))$-space. Remember that, for humans, it is assumed that $log(\mathcal{K}(\alpha)\epsilon)) = 0$.\footnote{One alternative approach for the results in \cref{fig:corrplot} would be to use the mean cooperation rates of algorithms obtained when selecting $\epsilon = 1/\mathcal{K}(\alpha)$, which yields a correction factor of zero in \cref{eq:klr}. However, this would require to use either low values of $\alpha$ or very low values of $\epsilon$. As pointed out at the end of \cref{sec:sim}, low value of $\alpha$ lead to anomalies when $d^{ic}$ is smaller than $50\%$ of its maximum possible distance, which is the case for many treatments in the meta data. Choosing larger values of $\alpha$ requires choosing values of $\epsilon$ below $1\%$, which were not considered in the simulation study.} \cref{sfig:corrplot} relies on the simulation results using optimistic initialization, the case of pessimistic initialization is covered in \cref{app:init_pes}.

\cref{sfig:correvo} shows the evolution of the Pearson correlation coefficients between the mean share of cooperative strategies observed for q-learning algorithms and the human cooperation rates in the first round as the experience of players increases with the number of games they played.  Until seven games, the number of treatments considered is constant. Thereafter, the number of treatments decreases. At $23$ games, only seven treatments remain. After peaking at around seven games, the Pearson correlation coefficient remains at a consistently high value above $0.8$. Overall, the analysis points towards a strong similarity between the conditions under which humans and q-learners cooperate.


\section{Conclusion}
\label{sec:conc}
This article demonstrates the usefulness of stochastic replicator dynamics in predicting the emergence of cooperation in the repeated prisoner's dilemma. Extensive simulations with $\epsilon$-greedy q-learning algorithms with one-period memory demonstrate that the potential function of the grim trigger strategy allows it to characterize the frontier between the parameter space dominated by defection and the parameter space inducive to cooperation.

Using meta data from laboratory experiments that analyze human choices in the repeated prisoner's dilemma, this article further explores how the results obtained from  simulations with q-learners extend to humans. Despite the limited variation in the prisoner's dilemma payoffs covered by the meta data, the analysis provides evidence that q-learners and humans learn to cooperate under similar conditions. 

This article provides guidance on how to select game parameters that would allow experiments with humans to discriminate between $sizeBAD $ and the $\mathcal{KLR}$. Future work could extend this paper's analysis to richer settings. The two-actions two-players scenario considered in this study likely oversimplifies many real world situations. How richer settings might affect the findings presented here remains to be studied.

\clearpage
\vspace{0.25cm}
\bibliography{coop_pd}

\newpage
\appendix
 \section{Stochastic Replicator Dynamics Under Grim Trigger, Proofs of Lemma \ref{lem:properties} and Proposition \ref{prop:klr_sg}} \label{app:lem}

\subsection*{Stochastic Replicator Dynamics Under Grim Trigger}
 
\cref{seq1:1} to \cref{seq1:4} provide the specifications and formulas needed to compute the potential function and its derivative under grim trigger. 

\begin{align}
	A &= \begin{bmatrix}
		r & -(1-\delta)s \\
		(1-\delta) & 0 
	\end{bmatrix} \label{seq1:1}\\[7.5pt]
	\mathbf{p} &= (p,1-p)\label{seq1:2}\\[7.5pt]
	U(p) &= -\int_0^p x\big((Ax)_i - x^{T}Ax\big)dx \label{seq1:3}\\[7.5pt]
	\frac{\partial U(p)}{\partial p} &=   - p \big((A\mathbf{p})_i - \mathbf{p} ^{T}A\mathbf{p} \big)\label{seq1:4}.
\end{align}

\noindent Using \cref{seq1:1} and \cref{seq1:2},  \cref{seq1:4} solves to

\begin{equation}\label{eq:ud}
	\frac{\partial U(p)}{\partial p} =   - p \big((r - (1 - \delta))(1-p)p- (1-\delta)s(1-p)^{2}\big).
\end{equation}

\noindent $p^{\star}$, as defined in property (i) of \cref{lem:properties}, is given by

\begin{equation}\label{eq:argmax}
	p^{\star} = \frac{(1-\delta)s}{r-(1-\delta)(1-s)}.
\end{equation}

The $sizeBAD$ measure \citep{dal2011evolution} is defined as the value of $p$ that solves 

\begin{equation}\label{eq:sb}
	p(r/(1-\delta)) + (1-p)(-s) = p.
\end{equation}

\subsection*{Proofs for \cref{lem:properties}}

The left-hand side of \cref{eq:sb} denotes the expected value of cooperating and playing grim trigger when the opponent randomizes between cooperating with probability $p$  and defecting with probability $(1-p)$. The right-hand side of the equation denotes the analogously defined expected value for defection. Solving \cref{eq:sb} for $p$ yields the same expression as \cref{eq:argmax}, which establishes property (i) of \cref{lem:properties}.

For property (ii), it can be verified that \cref{eq:argmax} is always in the interior of the interval $[0,1]$ and that the second derivative of \cref{seq1:3} is negative under the stated conditions. Additionally, we can rearrange \cref{eq:ud} to

\begin{equation}\label{eq:ud2}
	(1-\delta)s(1-\lambda),
\end{equation}

\noindent where $\lambda = p/p^{*}$. \cref{eq:ud2} is negative for $p>p^{*}$ and positive for $p<p^{*}$, which proves property (ii) of \cref{lem:properties}. 

Property (iii) can be established by verifying that \cref{eq:argmax} is either strictly larger than one, undefined, or negative if $r<(1-\delta)$. Additionally, we can rearrange \cref{eq:ud} to 

\begin{equation}\label{eq:ud1}
	(1-\delta)s(1-p^c) - p^c(1-\delta)(\frac{r}{1-\delta}-1).
\end{equation}

\noindent The first term in \cref{eq:ud1} is always weakly positive, while the second term is negative when $r<(1-\delta)$. This proves property (iii) of \cref{lem:properties}. Property (iv) can also be seen from \cref{eq:ud1} by noting that that $s=0$ and $r/(1-\delta) > 1$.  

For property (v), note that property (i) implies that $KE^{c} > 0$ and $KE^{d} > 0 $ under the stated conditions. Therefore, verifying property (v) amounts to verifying that $KE^{c}  - KE^{d}$ is strictly increasing in $r$ and strictly decreasing in $s$. $KE^{c} = U(p^{\star}) - U(1)$ and $KE^{d} = U(p^{\star}) - U(0) = U(p^{\star})$. Therefore, $KE^{c}  - KE^{d} = - U(1) = \frac{1}{12}(r-(1-\delta)(1+s))$.

\subsection*{Proof of \cref{prop:klr_sg}}

\begin{proof}
Fix a tuple $(\delta,r)$ such that $r > (1 - \delta)$. Furthermore, assume $s>0$. From properties (i) and (ii) of \cref{lem:properties}, we have that $sizeGOOD = 1 - p^{c}$, where $p^{c}$ is defined in \cref{eq:argmax}, which is increasing in $s$. From this, it follows that the supremum of $s$ consistent with $sizeGOOD \geq 0.5$ is given by $s = r/(1-\delta) - 1 \iff r = (1-\delta)(1+s)$. This implies that $KE^{c}-KE^{d} = -U(1) = 0 \iff KE^{c} = KE^{d}$ (see proof of property (v) of \cref{lem:properties}). From property (v) of \cref{lem:properties}, we know that the $\mathcal{KLR}$ is decreasing in $s$. The supremum of $s$ consistent with  $\mathcal{KLR} \geq 0$ can be found by solving $\mathcal{KLR} = 0 \iff KE^{c}/KE^{d} = \delta/(1-\delta)$. For $\delta > 0.5$, this implies that $KE^{c}>KE^{d}$. From property (v), this implies that the supremum of $s$, which is consistent with $\mathcal{KLR} \geq 0$, is smaller than the supremum of $s$, which is consistent with $sizeGOOD \geq 0$. This holds for any tuple $(\delta,r)$ that fulfills the stated conditions.
\end{proof}

\clearpage

 \section{Details on the Estimation of $\mathcal{K}(\alpha)$}\label{app:correction_factor}
 
To estimate the correction factor, the only observations used are those where the distance to the binding $IC$ constraint exceeds $0.35$. This corresponds to observations whose distance is weakly larger than $50\%$ of the maximal possible distance. When this condition is fulfilled, the $\mathcal{KLR}$ frontier provides a sharp boundary that facilitates the application of the procedure described next, which is carried out separately for each value of $\alpha$. 

For each tuple $(\delta, r)$, the value of the $\mathcal{KLR}$ for which we observe the maximum increase in the share of cooperative strategies is computed ($\mathcal{KLR}^{\star}$). Next, a grid of candidate values of $\mathcal{K}(\alpha)$ is defined. The grid consists of $1{,}000$ equally spaced points in the interval $[0.005,0.05]$. For each grid point, the candidate value for $\widehat{\mathcal{KLR}^{\star}} = \mathcal{K}(\alpha) \epsilon$ is computed. The grid point minimizing $MSE = \frac{1}{D}\sum_{d=1}^{D}(\widehat{\mathcal{KLR}^{\star}} - \mathcal{KLR}^{\star})^2$ is selected as the correction factor. $D$ denotes the number of distinct tuples $(\delta, r)$. \cref{tab:cal} reports the correction factor for each $\alpha$ for which the $MSE$ is minimized and the corresponding $MSE$.

\vspace{0.5cm}

\begin{table}[h!] 
\centering 
\caption{Estimated Correction Factors for Different Learning Rates $\alpha$} 
\label{tab:cal} 
\resizebox{0.95\textwidth}{!}{ 
\begin{threeparttable} 
{
\def\sym#1{\ifmmode^{#1}\else\(^{#1}\)\fi}
\sisetup{parse-numbers=false}
\setlength{\tabcolsep}{3pt} 
\begin{tabular}{ccccccccccc} 
\hline\hline   
& \multicolumn{10}{c}{{Learning Rate $\alpha$:}}  \Tstrut   \\
 & {$0.01$} & {$0.02$} & {$0.03$} & {$0.04$} & {$0.05$}  &  {$0.06$}  &  {$0.07$}  &  {$0.08$}  &  {$0.09$}  &  {$0.1$} \Bstrut \\ 
\hline 
$\mathcal{K}(\alpha)$  & $1/0.0320$ & $1/0.0152$  &  $1/0.008$ & $1/0.0051$  & $1/0.0030$  & $1/0.0022$& $1/0.0015$&  $1/0.0011$   & $1/0.0009$  & $1/0.0007$  \Tstrut \\ 
$MSE$ & $1.3103$ & $1.4515$ & $1.6578$ & $1.4095$ & $1.5300$ & $1.4267$ & $1.4347$ &  $1.8076$   & $1.3341$ & $1.4345$  \Tstrut \\ 
\hline\hline        
\end{tabular}
}
\begin{tablenotes} 
	\item
\end{tablenotes} 
\end{threeparttable} 
} 
\end{table} 
 
\clearpage

\section{Review of $\epsilon$-greedy Q-Learning}\label{app:ql}

Q-learning algorithms, originally proposed by \citet{watkins1989learning},  belong to a class of reinforcement learning algorithms and can be used to solve infinite-horizon Markov decision problems (MDP) by learning the value function of the MDP:

\begin{equation}\label{eq:valq}
	V(s) = \max_{a \in A}\{E[u_t|s,a] + \delta \sum_{s' \in S'} V(s) p(s'|s,a)\},
\end{equation}

\noindent where $a$ and $s$ denote the action and state in period $t$, $s'$ denotes the state in period $t+1$, $A$ denotes the action set from which the algorithm can select from in period $t$, $E[u_t|s,a]$ denotes the expected utility in period $t$ from choosing action $a$ in state $s$. $p(s'|s,a)$ is the time-invariant transition probability. Q-learners are off-policy algorithms that learn the value of each action-state combination, $Q(s,a)$, according to the following updating rule: 

\begin{equation}\label{eq:qupdt}
	Q_{t+1}(s,a) = Q_{t}(s,a) + \alpha \big(u_t(a,s) + \delta \max_{a \in A} Q_t(s,a) - Q_{t}(s,a) \big).
\end{equation}

In \cref{eq:qupdt}, $\alpha$ denotes the learning rate, which determines the speed with which the q-values are overwritten by novel payoff realizations. After the q-values have been learned, the value function can be retrieved by noting that $V(s) = \max_{a \in A} Q(s,a)$. \citet{watkins1992q} provide the first proof that, under fairly mild conditions, q-learning algorithms retrieve the optimal policy for MDPs.  

Q-learning algorithms learn the optimal value function by repeatedly applying the updating rule of \cref{eq:qupdt}. In $\epsilon$-greedy q-learning, in each period $t$, the action used to perform the update is selected uniformly at random with probability $\epsilon$, while the greedy action, that is, the action with the highest q-value in period $t$, is selected with probability $1-\epsilon$.

Multi-agent q-learning violates the MDP assumption of time-invariant transition probabilities. No theoretical convergence results for $MDPs$ are known in this case. Nevertheless, state-dependent multi-agent q-learning is an active area of research that typically relies on extensive Monte Carlo simulations to gain generalizeable insights \citep[see, for example,][]{calvano2020artificial}.  
 
 \clearpage
 
 \section{Pessimistic Initialization}\label{app:init_pes}
 
 For the results based on pessimistic initialization, note that the frontier still remains valid. The same correction factor as the one used for optimistic initialization is applied. Cooperation rates are lower throughout the entire parameter space considered. Note that the correlation coefficients between human and algorithmic cooperation rates remain high. While the qualitative results are unchanged in comparison to optimistic initialization, the cooperation rates based on optimistic initialization predict human cooperation rates better. 
 
 \begin{figure}[h!]
 	\centering
 	\begin{subfigure}[b]{0.48\textwidth}
 		\centering
 		\includegraphics[width=\textwidth]{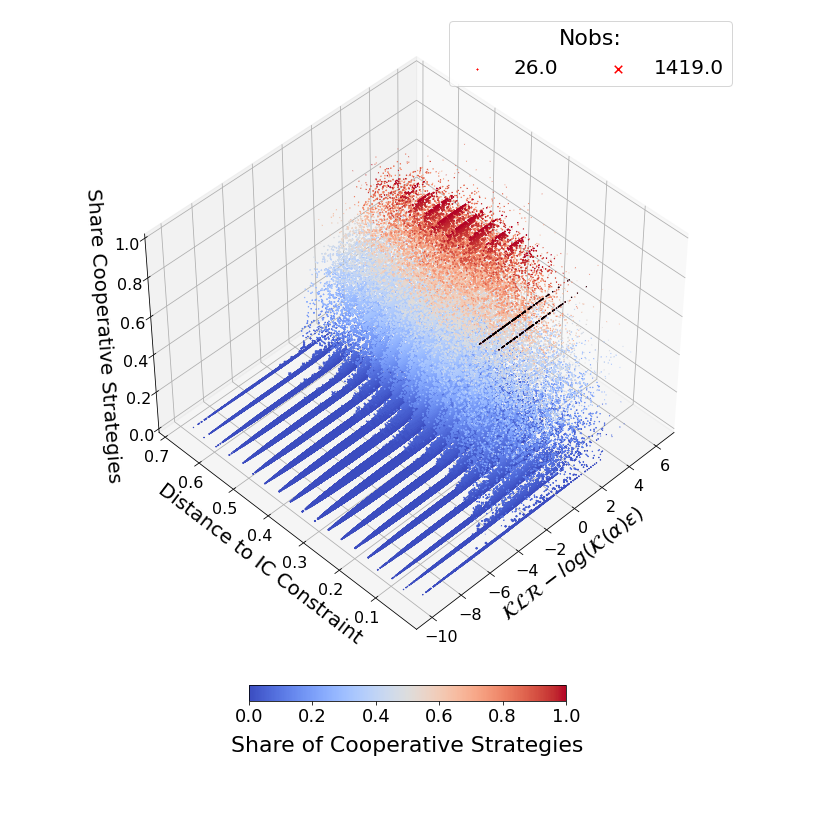}
 		\caption{3D Scatter -- All parameters}
 	\end{subfigure}
 	\hfill
 	\begin{subfigure}[b]{0.48\textwidth}
 		\centering
 		\includegraphics[width=\textwidth]{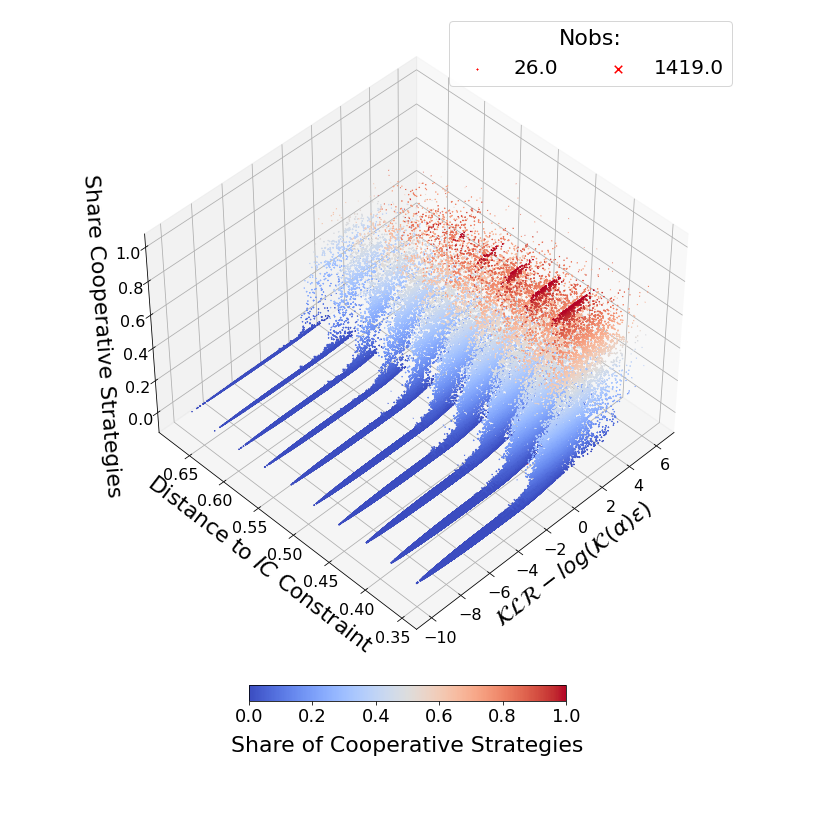}
 		\caption{3D Scatter -- Slack $IC$}
 	\end{subfigure}
 	\par\medskip
 	\par\medskip
 	\begin{subfigure}[b]{0.48\textwidth}
 		\centering
 		\includegraphics[width=\textwidth]{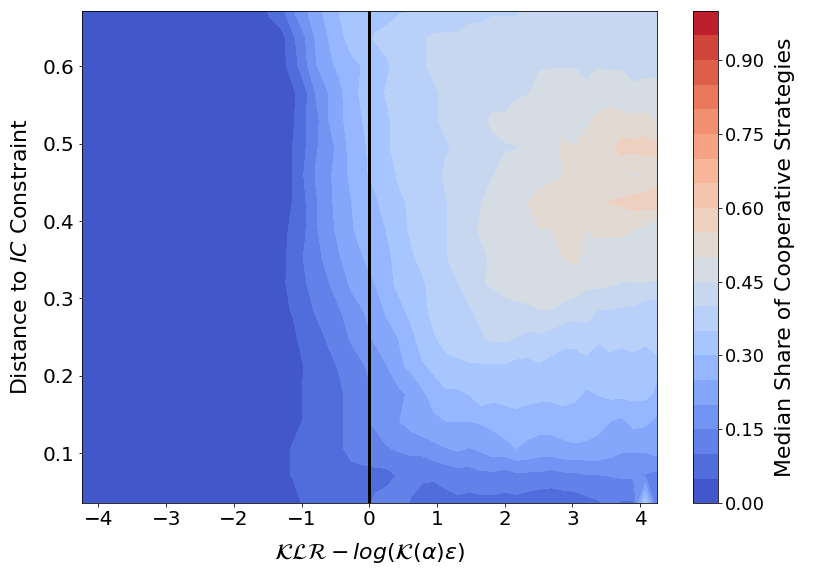}
 		\caption{Contour Plot -- All parameters}
 	\end{subfigure}
 	\hfill
 	\begin{subfigure}[b]{0.48\textwidth}
 		\centering
 		\includegraphics[width=\textwidth]{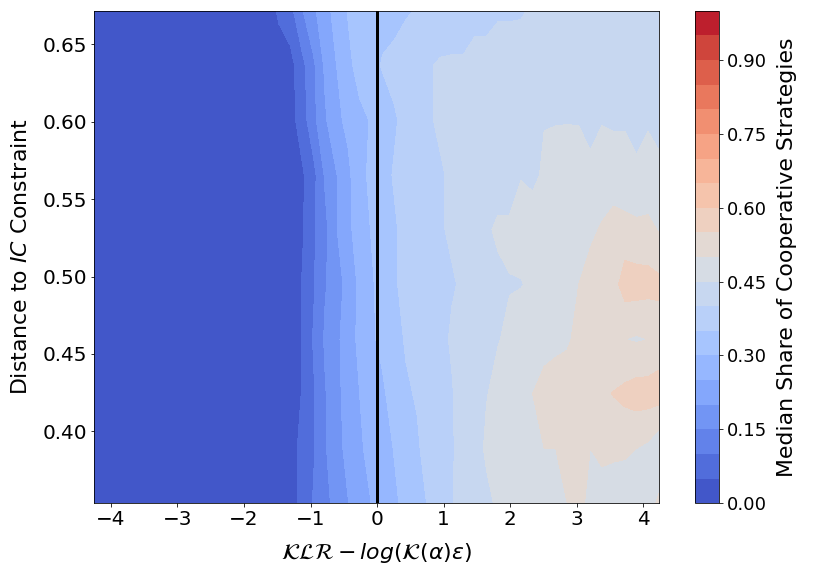}
 		\caption{Contour Plot -- Slack $IC$}
 	\end{subfigure}
 	\caption{Simulation Results -- Pessimistic Initialization}
	\caption*{\footnotesize Note: The size of each marker is calculated by first normalizing all variables to lie in the unit interval and then counting the number of neighboring observations in an open ball with radius $0.05$. Outliers with cooperation rates close to one near the binding $IC$ constraint are colored in gray. The contour plots show the isoquants for the median share of cooperative strategies computed in an open ball of radius $0.05$ around grid points of the normalized explanatory variables. The grid is based on $50$ equally spaced $\mathcal{KLR} - log(\mathcal{K}(\alpha)\epsilon)$ values for each distance to the binding $IC$ constraint.}
 \end{figure}
 
\begin{figure}[h!]
	\centering
	\begin{subfigure}[b]{0.45\textwidth}
		\centering
		\includegraphics[width=\textwidth]{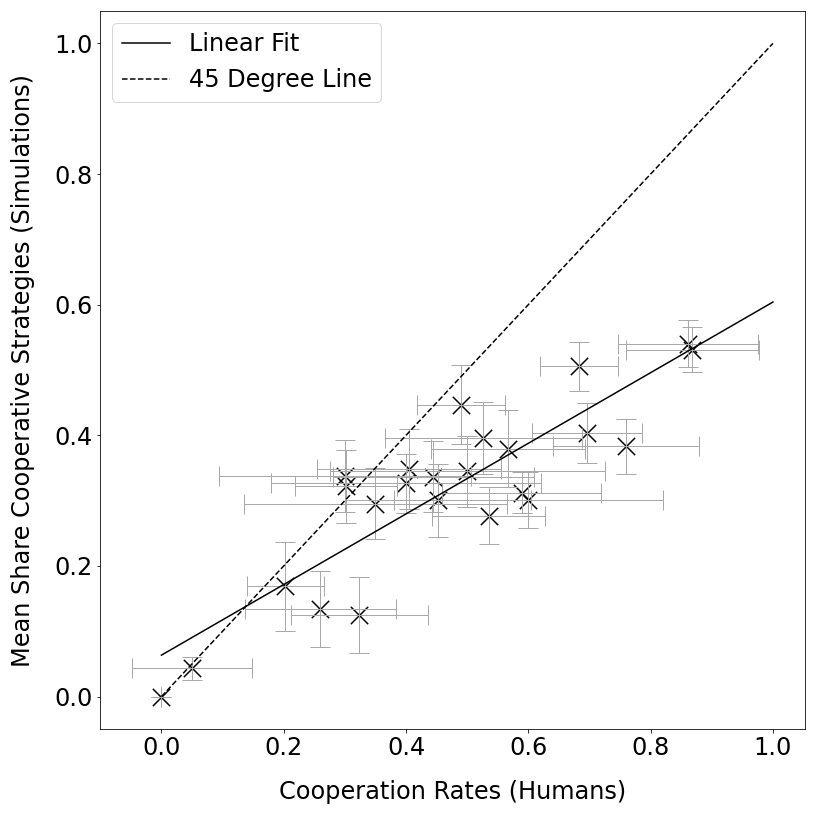}
		\caption{Relation Mean Cooperation Rates (Game 7)}
	\end{subfigure}
	\hfill
	\begin{subfigure}[b]{0.45\textwidth}
		\centering
		\includegraphics[width=\textwidth]{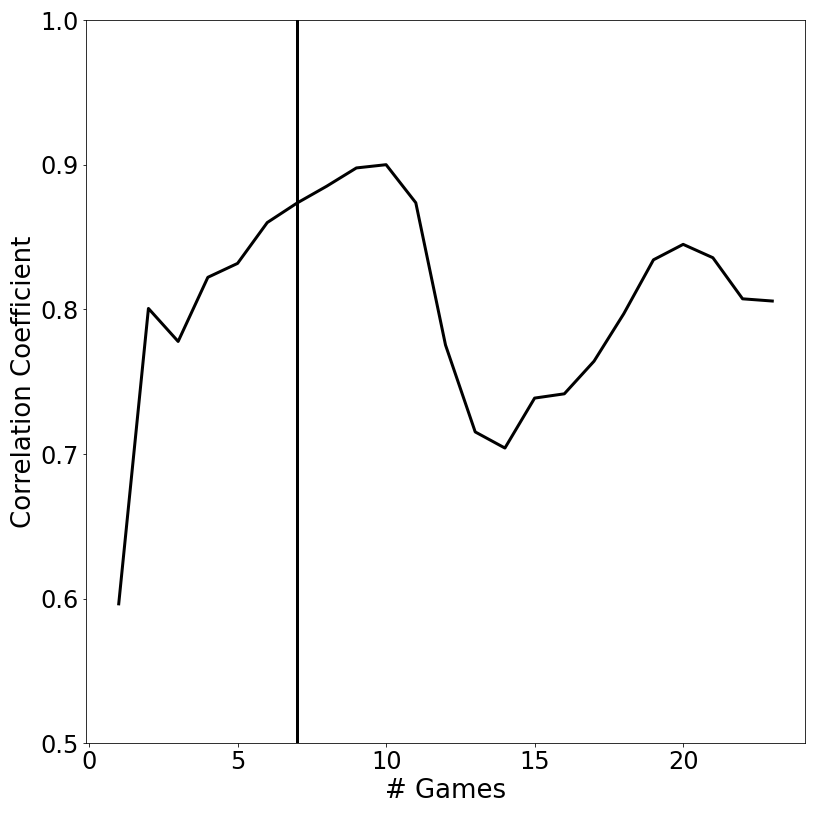}
		\caption{Evolution of Correlation Coefficient}
	\end{subfigure}
	\caption{Correlation Between Cooperation Rates Observed for Algorithms and Q-learners -- Pessimistic Initialization}
\end{figure}
  
\clearpage
 
\section{Results Based on Games Lasting Five Million }\label{app:5M}
  \begin{figure}[h!]
 	\centering
 	\begin{subfigure}[b]{0.46\textwidth}
 		\centering
 		\includegraphics[width=\textwidth]{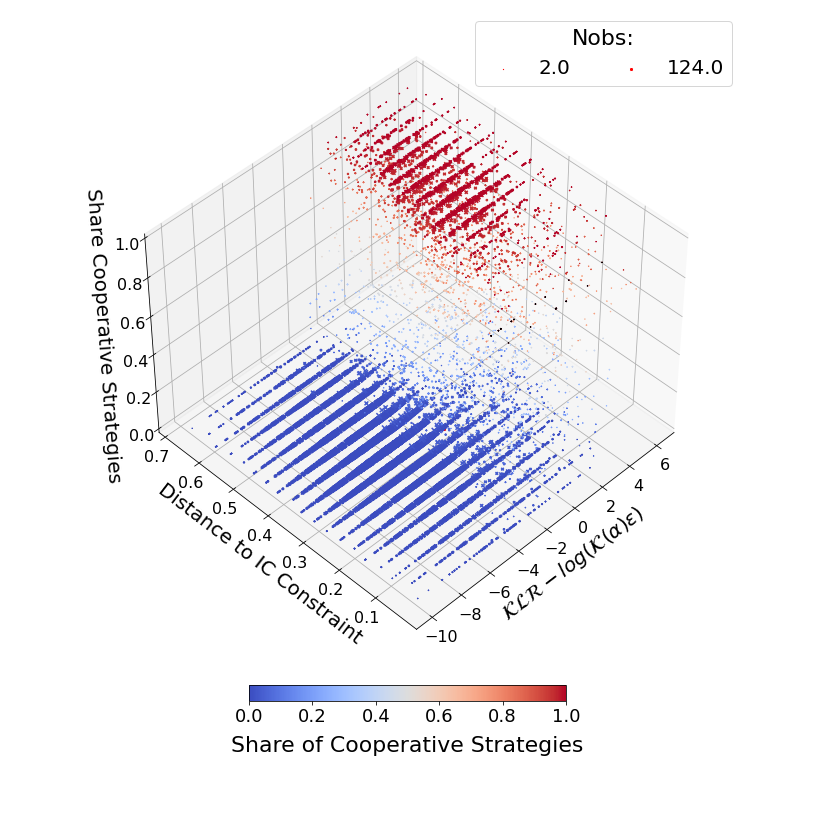}
 		\caption{3D Scatter -- All parameters}
 	\end{subfigure}
 	\hfill
 	\begin{subfigure}[b]{0.46\textwidth}
 		\centering
 		\includegraphics[width=\textwidth]{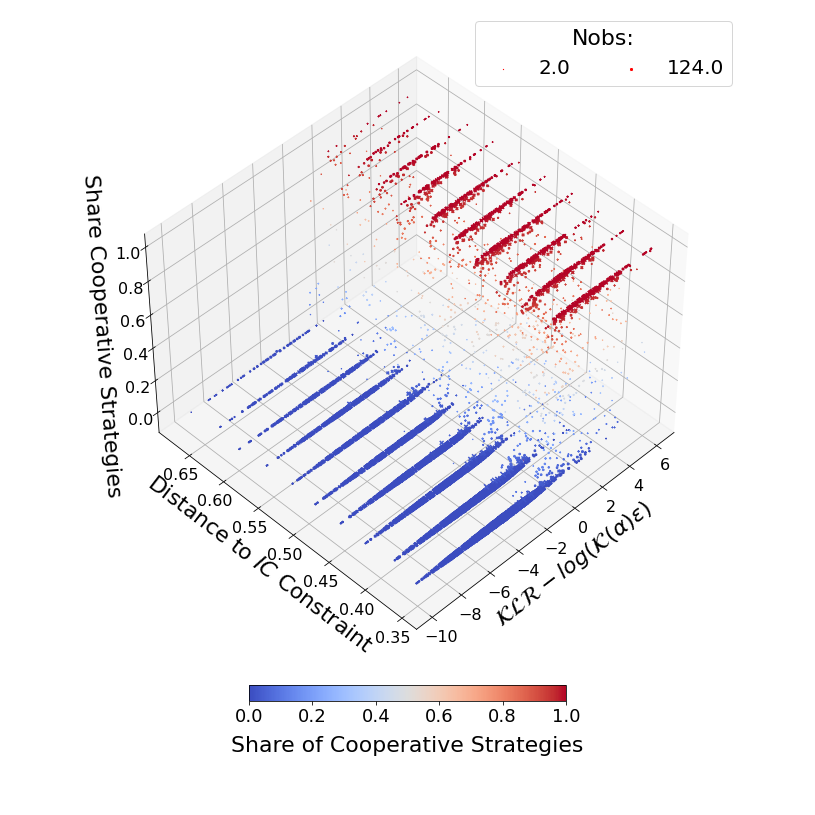}
 		\caption{3D Scatter -- Slack $IC$}
 	\end{subfigure}
 	\par\medskip
 	\par\medskip
 	\begin{subfigure}[b]{0.46\textwidth}
 		\centering
 		\includegraphics[width=\textwidth]{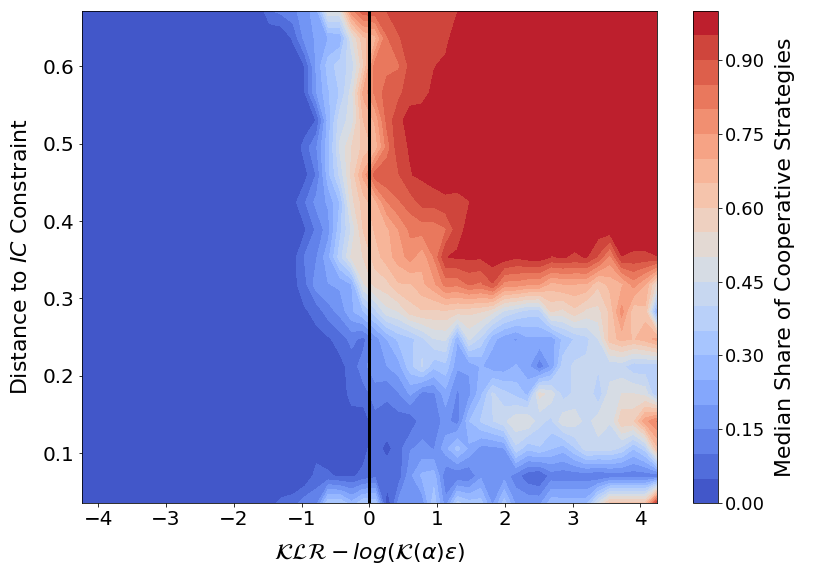}
 		\caption{Contour Plot -- All parameters}
 	\end{subfigure}
 	\hfill
 	\begin{subfigure}[b]{0.46\textwidth}
 		\centering
 		\includegraphics[width=\textwidth]{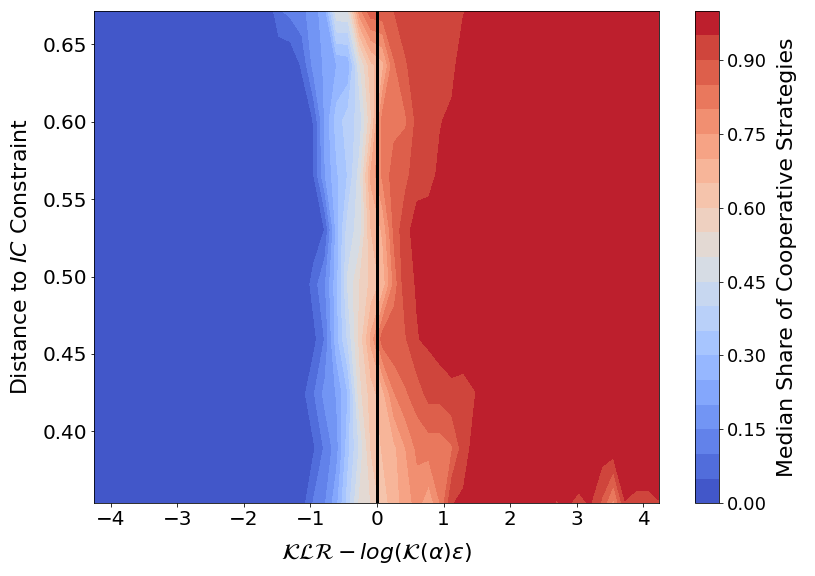}
 		\caption{Contour Plot -- Slack $IC$}
 	\end{subfigure}
 	\caption{Simulation Results -- Five Million Periods and Optimistic Initialization}
 	\caption*{\footnotesize Note: For $\epsilon$ and $\alpha$, the grid $[0.01,0.05,0.1]$ is used. The size of each marker is calculated by first normalizing all variables to lie in the unit interval and then counting the number of neighboring observations in an open ball with radius $0.05$. Outliers with cooperation rates close to one near the binding $IC$ constraint are colored in gray. The contour plots show the isoquants for the median share of cooperative strategies computed in an open ball of radius $0.05$ around grid points of the normalized explanatory variables. The grid is based on $50$ equally spaced $\mathcal{KLR} - log(\mathcal{K}(\alpha)\epsilon)$ values for each distance to the binding $IC$ constraint.}
 \end{figure}
 
\clearpage

\section{Decomposition of Non-Cooperative Strategy Profiles}\label{app:mean_het}

\begin{figure}[h!]
	\centering
	\begin{subfigure}[b]{0.46\textwidth}
		\centering
		\includegraphics[width=\textwidth]{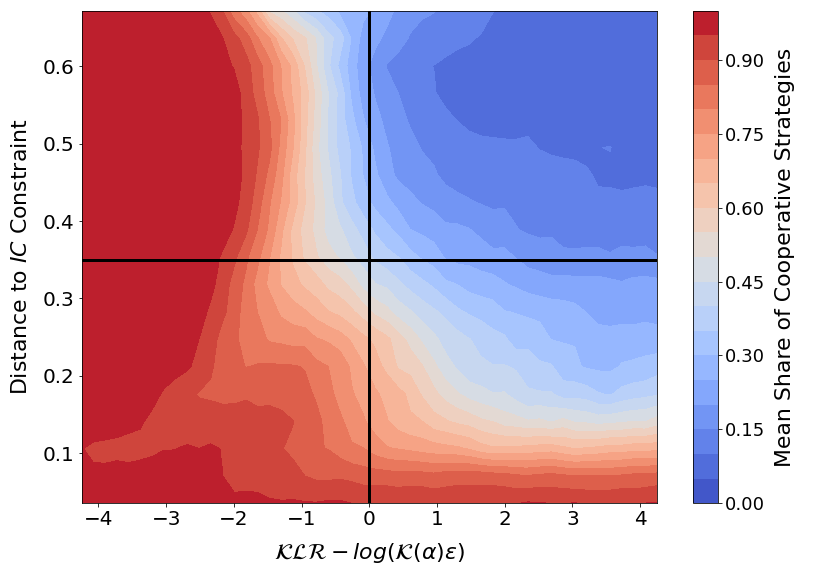}
		\caption{All Non-Cooperative Strategies}\label{sfig:mean_all_nc}
	\end{subfigure}
	\hfill
	\begin{subfigure}[b]{0.46\textwidth}
		\centering
		\includegraphics[width=\textwidth]{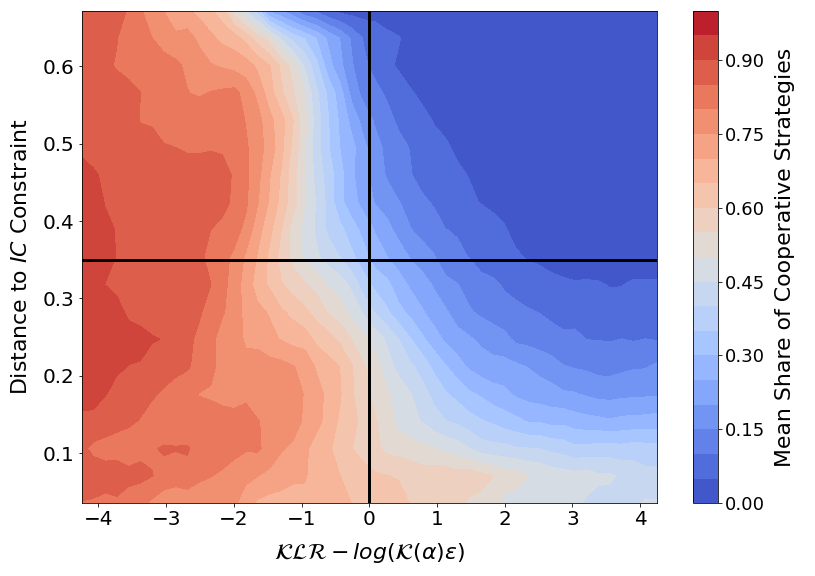}
		\caption{Mutual $ALLD$}\label{sfig:mean_alld}
	\end{subfigure}
	\par\medskip
	\par\medskip
	\begin{subfigure}[b]{0.46\textwidth}
		\centering
		\includegraphics[width=\textwidth]{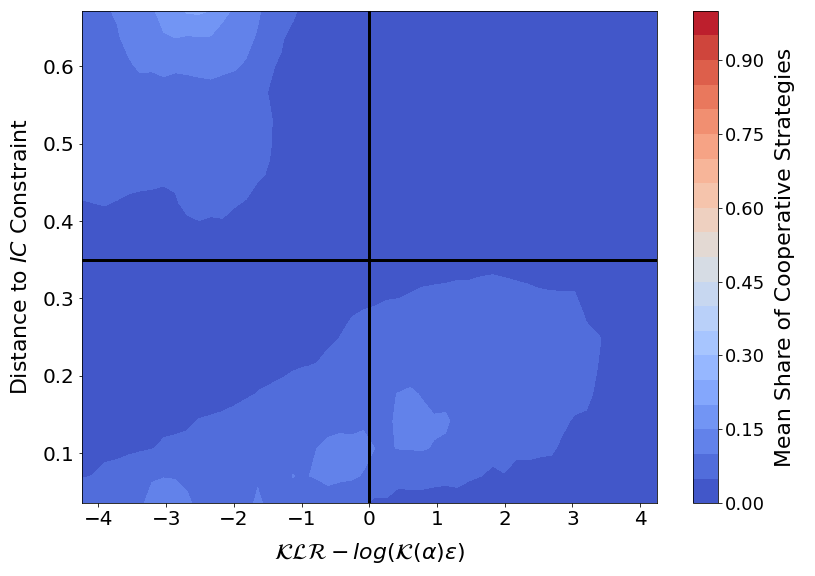}
		\caption{Mutual $GT$}\label{sfig:mean_gt}
	\end{subfigure}
	\hfill
	\begin{subfigure}[b]{0.46\textwidth}
		\centering
		\includegraphics[width=\textwidth]{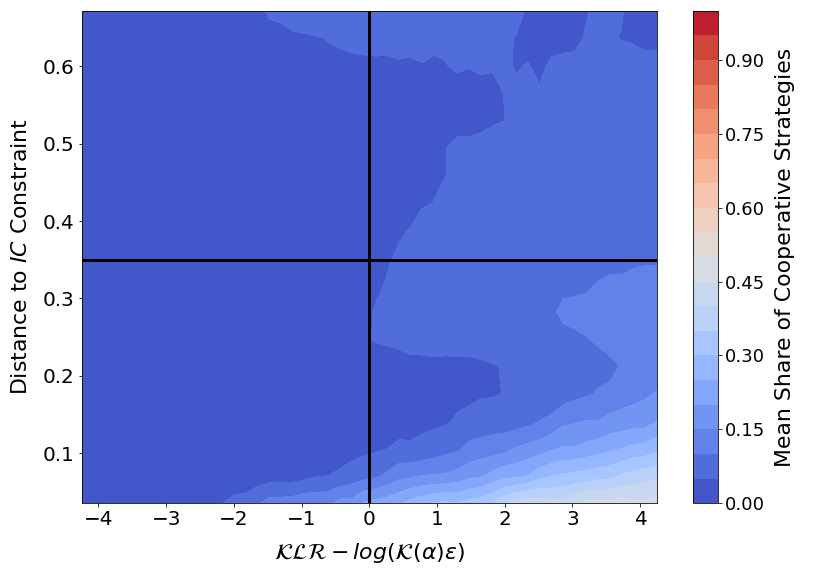}
		\caption{$EXPL$}\label{sfig:mean_expl}
	\end{subfigure}
	\caption{Simulation Results -- Decomposing Non-Cooperative Strategies}
\end{figure}

The \textit{mean} of the share of all non-cooperative strategy profiles is shown in \cref{sfig:mean_all_nc}. Non-cooperative strategy profiles mainly consist of mutual $ALLD$ (see \cref{sfig:mean_alld}), which specifies that both players always defect. As can be seen from \cref{sfig:mean_gt}, mutual grim trigger ($GT$) is present in a non-negligible manner to the left of the frontier, even as the distance to the $IC$ constraint is large. While grim trigger can sustain cooperation, it does not fulfill \cref{def:coop} and is therefore classified as non-cooperative. Exploitative strategies ($EXPL$) in which at least one player repeatedly incurs the sucker's payoff for at least $50\%$ of the time are particularly common close to the $IC$ constraint (see \cref{sfig:mean_expl}).  

\vspace{0.5cm}

\clearpage

\begin{landscape}
\section{Heterogeneity Analysis}\label{app:het}
\begin{figure}[h!]
	\centering
	\begin{subfigure}[b]{0.32\textwidth}
	\centering
	\includegraphics[width=\textwidth]{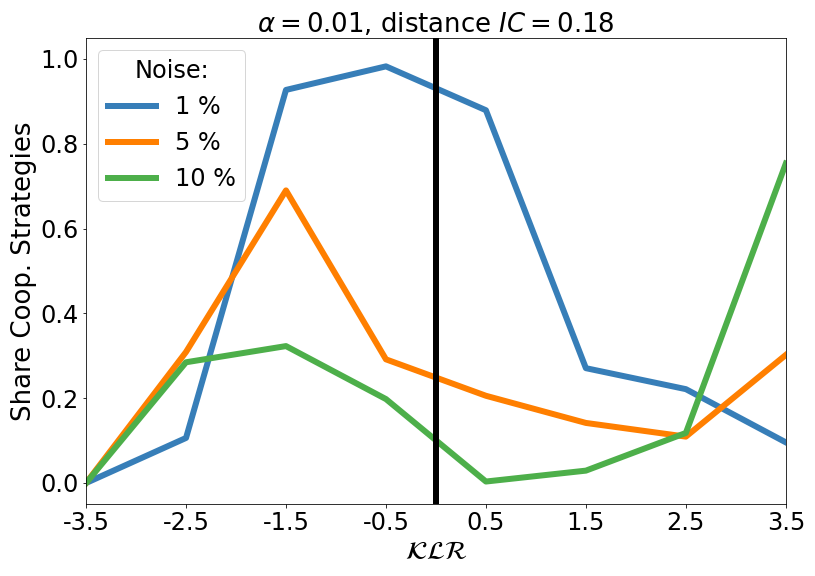}
	\end{subfigure}
	\hfill
	\begin{subfigure}[b]{0.32\textwidth}
	\centering
	\includegraphics[width=\textwidth]{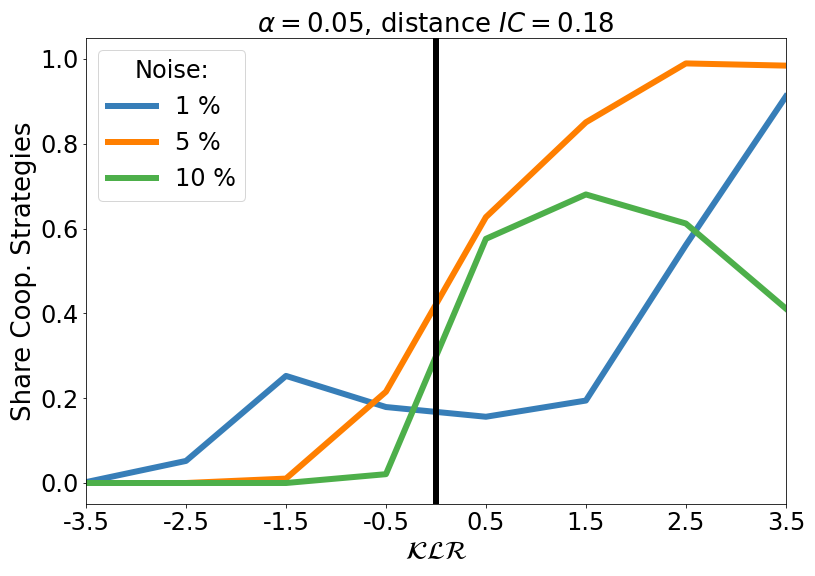}
	\end{subfigure}
	\hfill
	\begin{subfigure}[b]{0.32\textwidth}
	\centering
	\includegraphics[width=\textwidth]{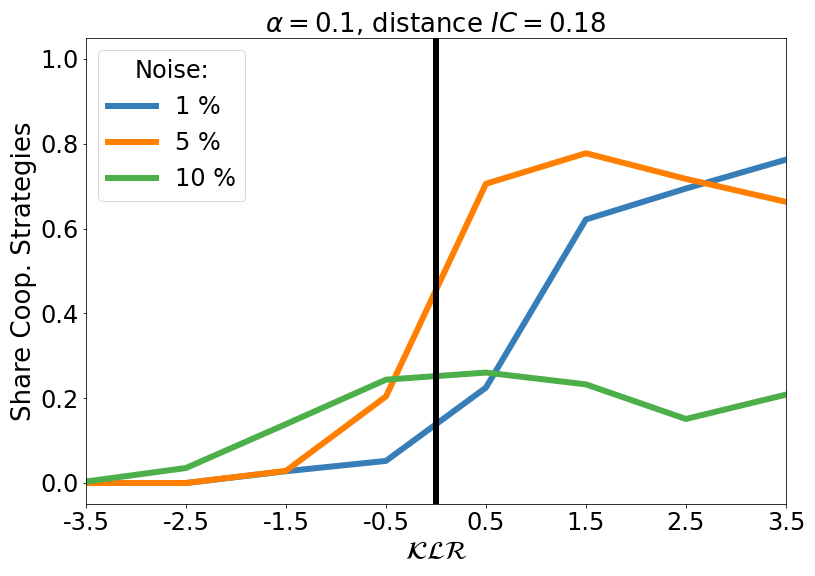}
	\end{subfigure}
	\vfill
	\begin{subfigure}[b]{0.32\textwidth}
	\centering
	\includegraphics[width=\textwidth]{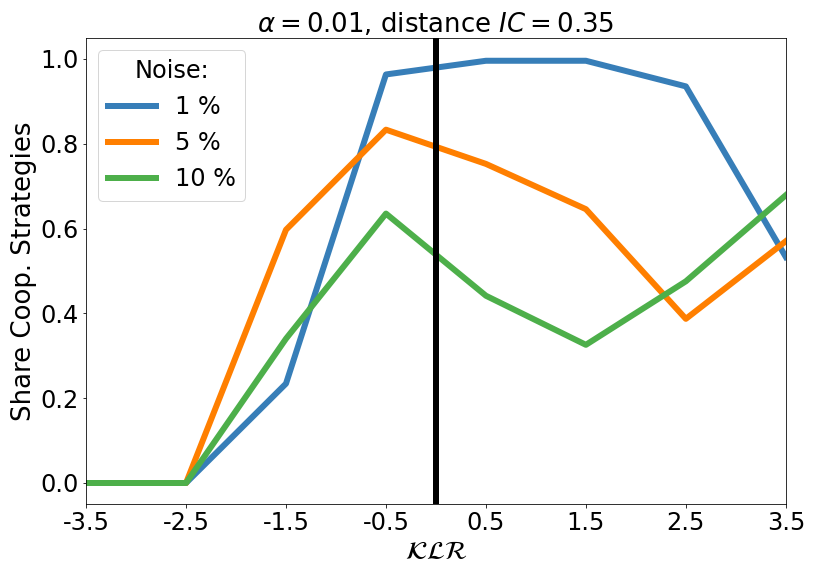}
	\end{subfigure}
	\hfill
	\begin{subfigure}[b]{0.32\textwidth}
	\centering
	\includegraphics[width=\textwidth]{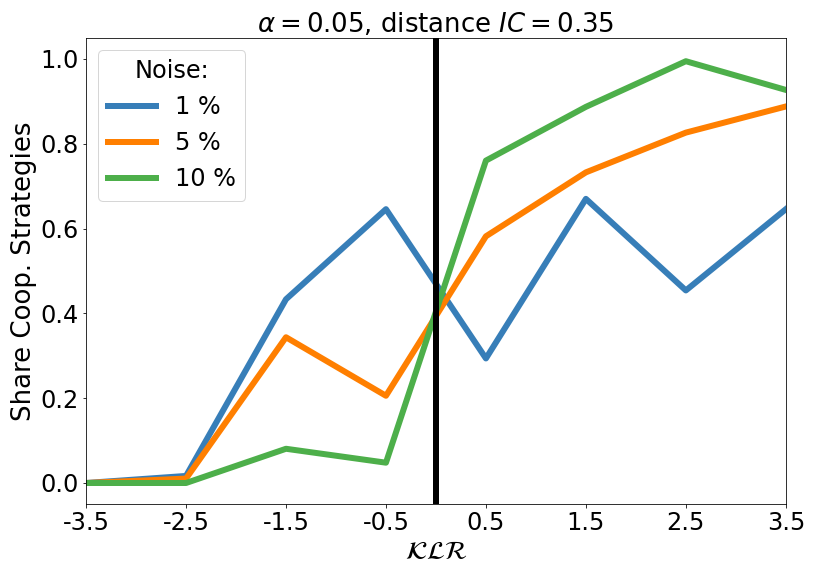}
	\end{subfigure}
	\hfill
	\begin{subfigure}[b]{0.32\textwidth}
	\centering
	\includegraphics[width=\textwidth]{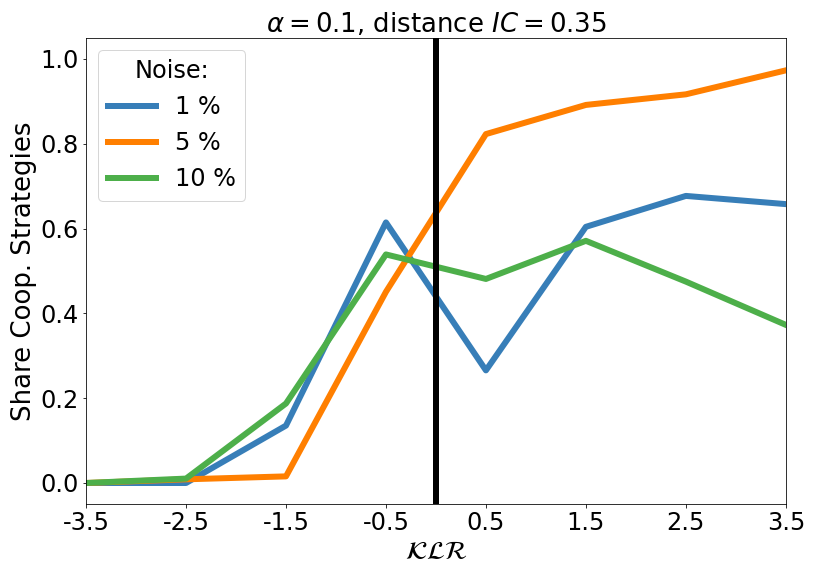}
	\end{subfigure}
	\vfill
	\begin{subfigure}[b]{0.32\textwidth}
	\centering
	\includegraphics[width=\textwidth]{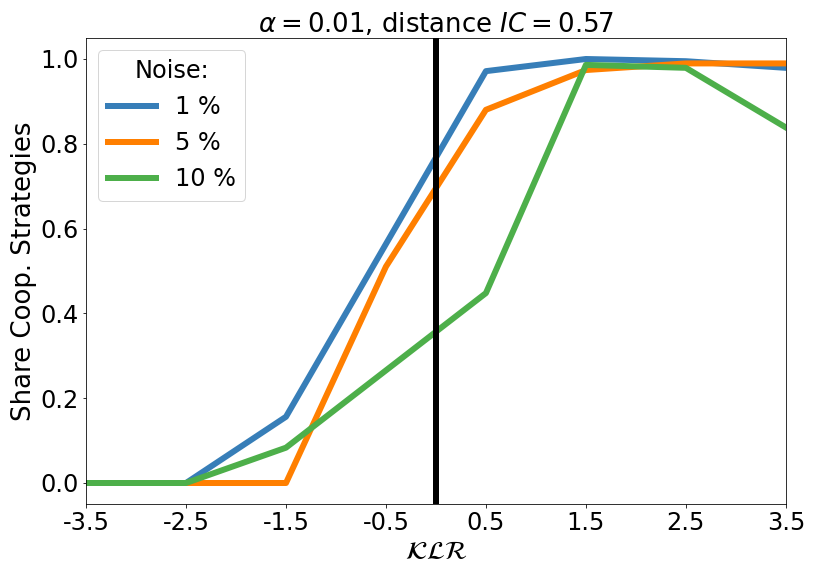}
	\end{subfigure}
	\hfill
	\begin{subfigure}[b]{0.32\textwidth}
	\centering
	\includegraphics[width=\textwidth]{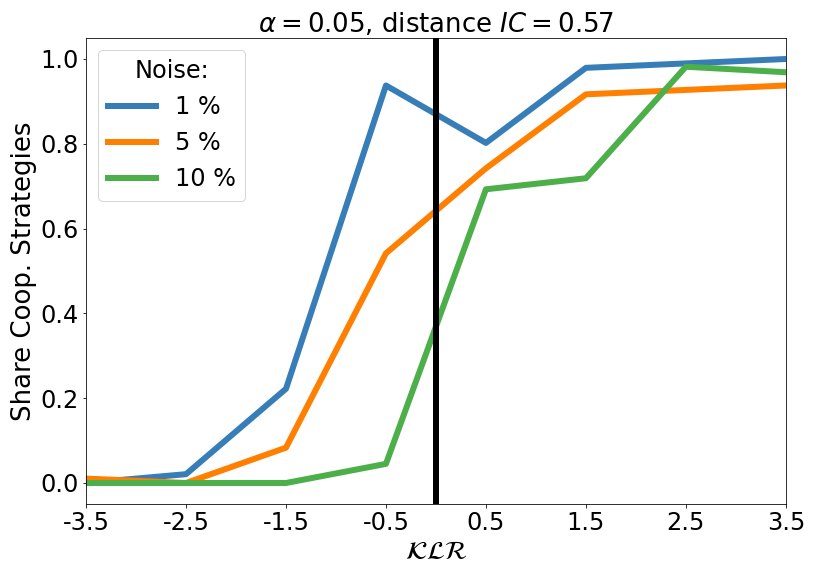}
	\end{subfigure}
	\hfill
	\begin{subfigure}[b]{0.32\textwidth}
	\centering
	\includegraphics[width=\textwidth]{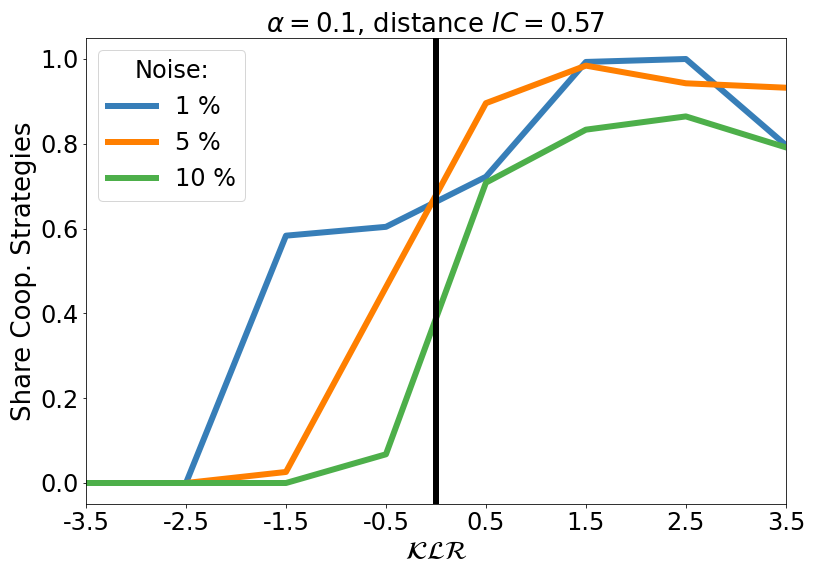}
	\end{subfigure}
	\caption{Average share of cooperative strategies for different combinations of $\alpha$, $\epsilon$ and $d^{ic}$.}
	\caption*{\footnotesize Note: Each row stands for a different distance to the binding $IC$ constraint, each column for a different parameter $\alpha$. The plots are obtained by computing the mean share of comparative strategies in bins defined by the adjacent integers $\mathcal{KLR} \in \{-4,4,3,\cdots,3,4\}$. The mid-points of the bins are used to create the plots.}
\end{figure}
\end{landscape}

\clearpage
\section{Summary Statistics Laboratory Experiments}\label{app:sum_stat}

\begin{table}[h!] 
	\centering 
	\caption{Summary Statistics Laboratory Experiments} 
	\label{tab:sum_stat} 
	\resizebox{0.95\textwidth}{!}{ 
		\begin{threeparttable} 
			{
				\def\sym#1{\ifmmode^{#1}\else\(^{#1}\)\fi}
				\sisetup{parse-numbers=false}
				\setlength{\tabcolsep}{10pt} 
				\begin{tabular}{lccccccc} 
					\hline\hline   
					& \multicolumn{6}{c}{{Experiment Characteristics:}}  \Tstrut   \\
				    Study & {$N$} & $\delta$ &{$r$} & {$s$} & {$\mathcal{KLR}$} & {$sizeGOOD$}  &  {$d^{ic}$}  \Bstrut \\ 
					\hline 
					\textbf{Dal B\'o (2005)} &  $42$ & $0.75$ & $0.46$  &  $0.38$ & $1.02$  & $0.69$   & $0.15$  \Tstrut \\ 
					& $60$  & $0.75$ & $0.55$ & $0.64$ & $0.56$ & $0.65$ & $0.21$   \Tstrut \\ 
					\textbf{Dreber et al. (2008)} $\dagger$  &  $22$ & $0.75$ & $0.5$  &  $0.5$ & $0.76$  & $0.67$   & $0.18$  \Tstrut \\ 
					\multicolumn{1}{c}{$\dagger$}  & $28$  & $0.75$ & $0.33$ & $0.67$ & $-2.95$ & $0.33$ & $0.059$   \Tstrut \\ 
					\textbf{Aoyagi and Frechette (2009)} &  $38$ & $0.9$ & $0.75$  &  $0.08$ & $10.20$  & $0.99$   & $0.46$  \Tstrut \\ 
					\textbf{Duffy and Ochs (2009)} &  $102$ & $0.9$ & $0.5$  &  $0.5$ & $3.51$  & $0.89$   & $0.28$  \Tstrut \\ 
					\textbf{Blonski et al. (2011)} $\dagger$ &  $20$ & $0.75$ & $0.5$  &  $0.5$ & $0.76$   & $0.67$   & $0.18$  \Tstrut \\ 
					&  $20$ & $0.75$ & $0.5$  &  $4$ & $-4.85$  & $0.2$   & $0.18$  \Tstrut \\ 
				    &  $20$ & $0.75$ & $0.55$  &  $0.27$ & $2.88$  & $0.81$   & $0.21$  \Tstrut \\ 
				    &  $20$ & $0.75$ & $0.57$  &  $0.71$ & $0.47$   & $0.64$   & $0.23$  \Tstrut \\ 
				    &  $20$ & $0.75$ & $0.67$  &  $2.33$ & $-2.00$   & $0.41$   & $0.29$  \Tstrut \\ 
				    &  $20$ & $0.875$ & $0.33$  &  $0.67$ & $0.51$   & $0.71$   & $0.15$  \Tstrut \\ 
				    &  $\mathbf{20}$ & $\mathbf{0.875}$ & $\mathbf{0.67}$  &  $\mathbf{2.33}$ & $\mathbf{-0.29}$   & $\mathbf{0.65}$   & $\mathbf{0.38}$  \Tstrut \\ 
				    \multicolumn{1}{c}{$\dagger$} &  $40$ & $0.75$ & $0.33$  &  $0.67$ & $-2.95$   & $0.33$   & $0.06$  \Tstrut \\ 
					\textbf{Dal B\'o and Frechette (2011a)} &  $38$ & $0.75$ & $0.6$  &  $0.52$ & $1.56$  & $0.73$   & $0.25$  \Tstrut \\ 
					\multicolumn{1}{c}{$\dagger$}  &  $44$ & $0.75$ & $0.28$  &  $0.52$ & $-5.07$  & $0.19$   & $0.02$   \Tstrut \\ 
					\multicolumn{1}{c}{$\dagger$}  &  $44$ & $0.75$ & $0.92$  &  $0.52$ & $3.32$   & $0.92$   & $0.44$   \Tstrut \\
					\multicolumn{1}{c}{$\dagger$}  &  $46$ & $0.5$ & $0.92$  &  $0.52$ & $1.28$   & $0.62$   & $0.30$   \Tstrut \\  
					&  $50$ & $0.5$ & $0.6$  &  $0.52$ & $-2.57$  & $0.28$   & $0.07$   \Tstrut \\ 
					\textbf{Bruttel and Kamecke (2012)} &  $36$ & $0.8$ & $0.46$  &  $0.38$ & $1.92$  & $0.77$   & $0.18$  \Tstrut \\ 
					\textbf{Fudenberg et al.(2012)} &  $36$ & $0.875$ & $0.75$  &  $0.25$ & $6.42$  & $0.95$   & $0.44$  \Tstrut \\ 
					\textbf{Kagel and Schley (2013)} &  $114$ & $0.75$ & $0.5$  &  $0.25$ & $2.65$  & $0.8$   & $0.18$  \Tstrut \\ 
					\textbf{Sherstyuk et al. (2013)} &  $56$ & $0.75$ & $0.5$  &  $0.125$ & $4.61$  & $0.89$   & $0.18$  \Tstrut \\ 
					\textbf{Frechette and Yuksel (2014)} &  $50$ & $0.75$ & $0.71$  &  $0.29$ & $4.02$  & $0.87$   & $0.33$  \Tstrut \\ 
					\textbf{Dal B\'o and Frechette (2015)} &  $20$ & $0.95$ & $0.28$  &  $0.52$ & $3.95$  & $0.89$   & $0.16$  \Tstrut \\ 
					\multicolumn{1}{c}{$\dagger$} &  $114$ & $0.75$  &  $0.28$  &  $0.52$ & $-5.07$  & $0.18$   & $0.02$  \Tstrut \\ 
					&  $116$  & $0.9$    &  $0.28$  &  $0.52$ & $1.16$  & $0.78$   & $0.13$  \Tstrut \\ 
					\multicolumn{1}{c}{$\dagger$}  &  $140$ & $0.5$    &  $0.92$  &  $0.52$ & $1.28$  & $0.62$   & $0.30$  \Tstrut \\ 
					\multicolumn{1}{c}{$\dagger$} &  $164$ & $0.75$  &  $0.92$  &  $0.52$ & $3.36$  & $0.84$   & $0.47$  \Tstrut \\ 
					\hline\hline        
				\end{tabular}
			}
			\begin{tablenotes} 
				\item Note: List of distinct treatments by study. The row marked in bold denotes the treatment for which the $\mathcal{KLR}<0$ and $sizeGOOD>0$. Only treatments used in the analysis are shown. $N$ refers to the number of observations observations in the first round of the seventh game. Note, some treatments in different studies are identical. In the analysis, duplicate treatments are aggregated. Treatments that occur twice across studies are indicated using a dagger symbol ($\dagger$) in the first column.
			\end{tablenotes} 
		\end{threeparttable} 
.	} 
\end{table} 

\clearpage

\section{Laboratory Experiments: Considering All Rounds}\label{app:all_rounds}

\begin{figure}[h!]
	\centering
	\includegraphics[width=0.65\textwidth]{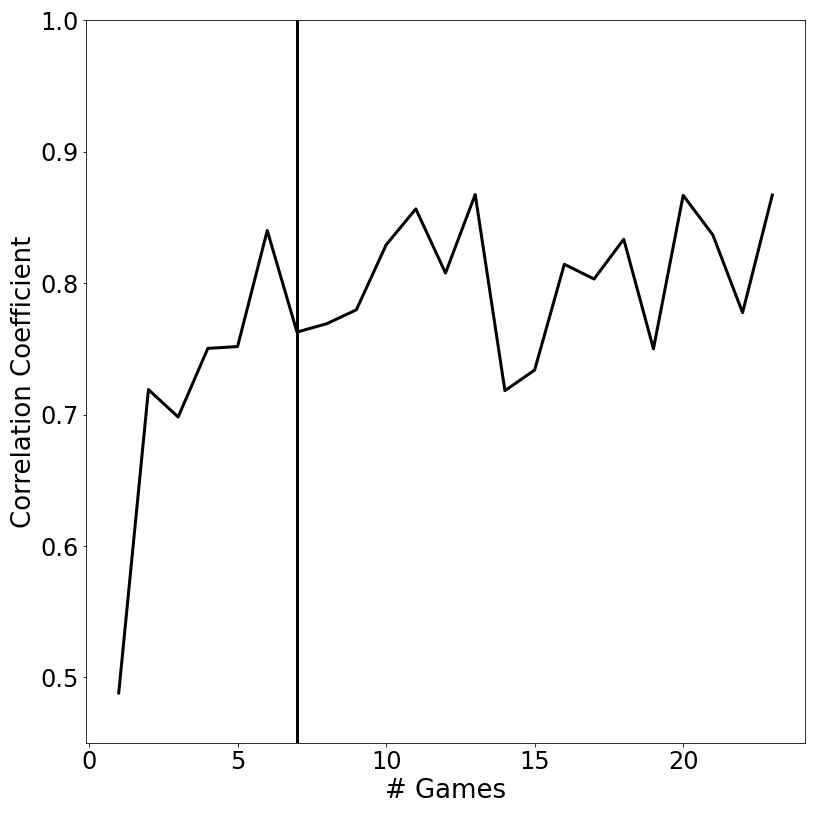}
	\caption{Evolution of Correlation Between Cooperation Rates Observed for Algorithms and Q-learners -- Using Choices Observed Across All Rounds}
\end{figure}

\end{document}